\documentclass[prd,showkeys,floatfix,twocolumn,amsmath,amssymb,floatfix]{revtex4}
\usepackage{graphicx}
\usepackage{epstopdf}
\usepackage{multirow}
\usepackage{subfigure}
\usepackage[colorlinks,citecolor=blue,anchorcolor=red,menucolor=red,linkcolor=red,filecolor=red,runcolor=red,urlcolor=blue,frenchlinks=red]{hyperref}
\newcommand{\feynp}[1]{#1\kern-0.45em/}
\allowdisplaybreaks[3]
\begin{document}

\title{Radiative decays of $P$-wave charmed baryons in the $SU(3)$ flavor $\bf6_F$ representation}

\author{Xuan Luo$^1$}
\email{cnxluo@seu.edu.cn}
\author{Hua-Xing Chen$^1$}
\email{hxchen@seu.edu.cn}
\author{Er-Liang Cui$^2$}
\email{erliang.cui@nwafu.edu.cn}
\author{Hui-Min Yang$^3$}
\email{hmyang@pku.edu.cn}
\author{Dan Zhou$^4$}
\email{danzhou@hebtu.edu.cn}
\author{Zhi-Yong Zhou$^1$}
\email{zhouzhy@seu.edu.cn}
\affiliation{$^1$School of Physics, Southeast University, Nanjing 210094, China
\\
$^2$College of Science, Northwest A\&F University, Yangling 712100, China
\\
$^3$School of Physics and Center of High Energy Physics, Peking University, Beijing 100871, China
\\
$^4$Department of Physics and Hebei Key Laboratory of Photophysics Research and Application, Hebei Normal University, Shijiazhuang 050024, China}

\begin{abstract}
We perform a comprehensive investigation of the radiative decays of $P$-wave charmed baryons in the $SU(3)$ flavor $\mathbf{6}_F$ representation, employing the light-cone QCD sum rule approach within the framework of heavy quark effective theory. We analyze their electromagnetic transitions into ground-state charmed baryons via photon emission. When combined with the mass spectra and strong decay properties previously studied in Ref.~\cite{Yang:2021lce}, our results constitute a systematic and complete QCD sum rule analysis of the $P$-wave singly charmed baryons within the framework of heavy quark effective theory. As summarized in Table~\ref{tab:result}, several excited charmed baryons are found to possess suppressed strong decay widths, thereby rendering their radiative decay channels particularly significant for experimental identification and theoretical understanding.
\end{abstract}
\pagenumbering{arabic}
\maketitle

\section{Introduction}

Over the past few decades, significant experimental progress has been achieved in the observation of excited charmed baryons. For recent developments in this field, we refer the reader to Refs.~\cite{Crede:2013kia,Chen:2016spr,Cheng:2021qpd,Chen:2022asf}. In particular, the LHCb Collaboration has reported the simultaneous observation of five excited $\Omega_c$ states in the $\Xi_c K^-$ invariant mass spectrum~\cite{LHCb:2017uwr}, identified as $\Omega_c(3000)^0$, $\Omega_c(3050)^0$, $\Omega_c(3066)^0$, $\Omega_c(3090)^0$, and $\Omega_c(3119)^0$. Some of these states were later confirmed by Belle~\cite{Yelton:2017qxg} and through subsequent analyses by LHCb~\cite{LHCb:2021ptx}. According to the most recent data compiled by the Particle Data Group~\cite{PDG2024}, their measured masses and decay widths are given as follows:
\begin{eqnarray}
\Omega_c(3000)^0 : \rm M &=&3000.4 \pm 0.2 \pm 0.1^{+0.3}
_{-0.5}{\rm~MeV} \, ,
 \nonumber
\\       \Gamma &=&4.5 \pm 0.6 \pm 0.3~\rm MeV\, ;
\\ \Omega_c(3050)^0 :\rm M &=& 3050.2 \pm 0.1 \pm 0.1^{+0.3}
_{-0.5}{\rm~MeV} \, ,
\nonumber
\\      \Gamma&=& 0.8 \pm 0.2 \pm 0.1{\rm~MeV} \, ;
\\ \Omega_c(3066)^0 :\rm M &=& 3065.6 \pm 0.1 \pm 0.3^{+0.3}
_{-0.5}{\rm~MeV} \, ,
\nonumber
\\       \Gamma&=&3.5 \pm 0.4 \pm 0.2{\rm~MeV} \, ;
\\ \Omega_c(3090)^0 :\rm M &=&3090.2 \pm 0.3 \pm 0.5^{+0.3}
_{-0.5}{\rm~MeV} \, ,
\nonumber
\\ \Gamma&=& 8.7 \pm 1.0 \pm 0.8{\rm~MeV} \, ;
\\ \Omega_c(3119)^0 :\rm M &=& 3119.1 \pm0.3 \pm 0.9^{+0.3}
_{-0.5}{\rm~MeV} \, ,
\nonumber
\\       \Gamma&=& 1.1 \pm 0.8 \pm 0.4{\rm~MeV} \, .
\end{eqnarray}
These excited $\Omega_c$ states are promising candidates for $P$-wave charmed baryons belonging to the $SU(3)$ flavor $\mathbf{6}_F$ representation. In addition to them, several other excited charmed baryons observed in experiments can also be interpreted as $P$-wave states in the same flavor representation. These include the following:
\begin{itemize}

\item In 2005 the Belle Collaboration observed an excited charmed baryon $\Sigma_c$, referred to as $\Sigma_c(2800)$, in the $\Lambda_c \pi$ decay channel~\cite{CLEO:2000mbh}. The measured masses and decay widths are~\cite{PDG2024}:
\begin{eqnarray}
\Sigma_c(2800)^0 : \rm M &=&2806^{+ 5}_{- 7}{\rm~MeV} \, ,
 \nonumber
\\       \Gamma &=&72^{+22}_{-15}~\rm MeV\, ;
\\ \Sigma_c(2800)^+ :\rm M &=& 2792^{+ 14}_{- 5}{\rm~MeV} \, ,
\nonumber
\\      \Gamma &=&62^{+37~+52}_{-23~-38}{\rm~MeV} \, .
\\ \Sigma_c(2800)^{++} :\rm M &=& 2801^{+ 4}_{-6}{\rm~MeV} \, ,
\nonumber
\\     \Gamma &=&75^{+18~+12}_{-13~-11}{\rm~MeV}\, .
\end{eqnarray}

\item In 2020 the LHCb Collaboration reported the observation of three excited $\Xi_c$ states, denoted as $\Xi_c(2923)^0$, $\Xi_c(2939)^0$, and $\Xi_c(2965)^0$, in the $\Lambda_c^+ K^-$ invariant mass spectrum~\cite{LHCb:2020iby}. Their measured properties are~\cite{PDG2024}:
\begin{eqnarray}
    \nonumber              \Xi_{c}(2923)^{0}:M&=&2923.04 \pm 0.25 \pm 0.20 \pm 0.14 \mbox{ MeV} \, ,
    \\                         \Gamma&=&7.1 \pm 0.8 \pm 1.8 \mbox{ MeV} \, ,
    \\ \nonumber        \Xi_{c}(2939)^{0}:M&=&2938.55 \pm 0.21 \pm 0.17 \pm 0.14 \mbox{ MeV} \, ,
    \\                         \Gamma&=&10.2 \pm 0.8 \pm 1.1 \mbox{ MeV} \, ,
    \\ \nonumber        \Xi_{c}(2965)^{0}:M&=&2964.88 \pm 0.26 \pm 0.14 \pm 0.14 \mbox{ MeV} \, ,
    \\                                 \Gamma&=&14.1 \pm 0.9 \pm 1.3 \mbox{ MeV} \, .
\end{eqnarray}
\end{itemize}

Investigations into the mass spectra, as well as the production and decay properties, are essential for understanding the internal structure and underlying dynamics of excited charmed baryons. To this end, a wide range of theoretical approaches and phenomenological models have been developed, including various quark models~\cite{Capstick:1986bm,Ebert:2007nw,Garcilazo:2007eh,Roberts:2007ni,Zhong:2007gp,Valcarce:2008dr,Ebert:2011kk,Ortega:2012cx,Yoshida:2015tia,Nagahiro:2016nsx,Wang:2017kfr,Lu:2020ivo,Chen:2021eyk}, lattice QCD simulations~\cite{Bowler:1996ws,Burch:2008qx,Brown:2014ena,Padmanath:2013bla,Padmanath:2017lng,Bahtiyar:2020uuj}, QCD sum rules~\cite{Chen:2015kpa,Bagan:1991sg,Neubert:1991sp,Broadhurst:1991fc,Huang:1994zj,Dai:1996yw,Groote:1996em,Colangelo:1998ga,Huang:2000tn,Mao:2015gya,
Zhu:2000py,Lee:2000tb,Wang:2003zp,Duraes:2007te,Liu:2007fg,Zhang:2008pm,Wang:2010it,Zhou:2014ytp,Zhou:2015ywa,Wang:2017zjw,Aliev:2018ube,Xu:2020ofp,Chen:2017sci,Agaev:2017lip,Agaev:2017nn}, the $^3P_0$ decay model~\cite{Chen:2007xf,Chen:2017gnu,Zhao:2017fov,Ye:2017yvl}, chiral perturbation theory~\cite{Lu:2014ina,Cheng:2015naa,Cheng:2006dk,Cho:1994vg,Pirjol:1997nh,Chiladze:1997ev,Blechman:2003mq,Huang:1995ke}, the light-front quark model~\cite{Tawfiq:1998nk,Tawfiq:1999vz}, relativistic three-quark models~\cite{Ivanov:1999bk,Ivanov:1998qe,Korner:1994nh,Hussain:1999sp}, nonrelativistic quark models~\cite{Albertus:2005zy}, the MIT bag model~\cite{Hwang:2006df}, and the Bethe–Salpeter formalism~\cite{Guo:2007qu}, among others.

We have systematically investigated the properties of $P$-wave charmed and bottom baryons using QCD sum rules~\cite{Gimenez:2005nt,Shifman:1978bx,Shifman:1978by,Reinders:1984sr,Narison:2002woh,Nielsen:2009uh,Gubler:2018ctz} and light-cone sum rules~\cite{Braun:1988qv,Chernyak:1990ag,Ball:1998je,Ball:2006wn,Ball:2004rg,Ball:1998kk,Ball:1998sk,Ball:1998ff,Ball:2007rt,Ball:2007zt,Ball:2002ps} within the framework of heavy quark effective theory (HQET)~\cite{Eichten:1989zv,Grinstein:1990mj,Falk:1990yz}. Our previous works include:
\begin{itemize}

\item In Ref.~\cite{Yang:2022oog,Tan:2023opd,Luo:2024jov,Wang:2024rai} we carried out detailed analyses of the mass spectra, as well as the strong and radiative decay properties, of $P$-wave singly bottom baryons belonging to the $SU(3)$ flavor $\mathbf{\bar 3}_F$ representation.

\item In Ref.~\cite{Yang:2019cvw,Cui:2019dzj,Yang:2020zrh,Chen:2020mpy,Luo:2024jov} we carried out detailed analyses of the mass spectra, as well as the strong and radiative decay properties, of $P$-wave singly bottom baryons belonging to the $SU(3)$ flavor $\mathbf{6}_F$ representation.

\item In Ref.~\cite{Chen:2015kpa,Chen:2017sci,Yang:2023fsc,Luo:2025jpn} we carried out detailed analyses of the mass spectra, as well as the strong and radiative decay properties, of $P$-wave singly charmed baryons belonging to the $SU(3)$ flavor $\mathbf{\bar 3}_F$ representation.

\item In Ref.~\cite{Chen:2015kpa,Chen:2017sci,Yang:2020zjl,Yang:2021lce} we carried out detailed analyses of the mass spectra, as well as the strong  decay properties, of $P$-wave singly charmed baryons belonging to the $SU(3)$ flavor $\mathbf{6}_F$ representation.

\end{itemize}
Despite this progress, a key aspect of the spectroscopy of $P$-wave singly heavy baryons remains insufficiently understood:
\begin{itemize}

\item The radiative decays of $P$-wave singly charmed baryons in the $SU(3)$ flavor $\mathbf{6}_F$ representation have not yet been systematically studied.

\end{itemize}
The present work is devoted to filling this important gap.

Compared to strong decays, radiative decays also play a crucial role, particularly in scenarios where the mass difference between the initial and final states is too small to permit strong decays. Experimentally, radiative transitions such as $\Xi_c(2815)/\Xi_c(2790) \to \Xi_c \gamma$ have already been observed by the Belle Collaboration~\cite{Belle:2020ozq}. Moreover, further radiative transitions are expected to be measured at upcoming experiments, including Belle-II and LHCb. On the theoretical side, a variety of models have been employed to study radiative decays of heavy baryons~\cite{Cheng:1992xi,Wang:2009ic,Tawfiq:1999cf,Gamermann:2010ga,Wang:2009cd,Jiang:2015xqa,Zhu:1998ih,Aliev:2014bma,Aliev:2009jt,Aliev:2016xvq,Aliev:2011bm,Chow:1995nw,Bahtiyar:2016dom,Bahtiyar:2015sga,Ivanov:1998wj,Savage:1994wa,Banuls:1999br,Ortiz-Pacheco:2023kjn}. However, the majority of these works focus on the radiative decays of $S$-wave singly heavy baryons. In contrast, only a limited number of studies have explored radiative transitions involving $P$-wave heavy baryons~\cite{Cho:1994vg,Ivanov:1999bk,Ivanov:1998wj,Zhu:2000py,Tawfiq:1999cf,Gamermann:2010ga,Chow:1995nw}. In this work, we focus on this less-explored area, aiming to deepen the theoretical understanding of radiative decays of $P$-wave singly heavy baryons.

In Ref.~\cite{Luo:2024jov} we investigated the radiative decays of $P$-wave bottom baryons in the $SU(3)$ flavor $\mathbf{6}_F$ representation using light-cone QCD sum rules. In the present study, we extend this analysis by replacing the bottom quark with a charm quark, thereby performing a comprehensive investigation of the radiative decay properties of $P$-wave singly charmed baryons. Moreover, we incorporate the mixing effects between different HQET multiplets, which were neglected in our previous analysis of $P$-wave bottom baryons~\cite{Luo:2024jov}, but were taken into account in our earlier study of $P$-wave charmed baryons within the same $\mathbf{6}_F$ flavor representation~\cite{Yang:2021lce}. This refinement leads to a more complete and accurate description of the relevant radiative transitions.

The remainder of this paper is organized as follows. In Sec.~\ref{secpcharmed} we briefly introduce the notations and theoretical framework for describing $P$-wave charmed baryons. Sec.~\ref{sec:6FD} is devoted to a detailed analysis of the radiative decay properties of $P$-wave charmed baryons in the $SU(3)$ flavor $\mathbf{6}_F$ representation. Finally, we summarize our results and present concluding remarks in Sec.~\ref{secsummry}.

\section{Notations and Mass spectra}
\label{secpcharmed}

\begin{figure*}[hbtp]
\begin{center}
\includegraphics[width=0.95\textwidth]{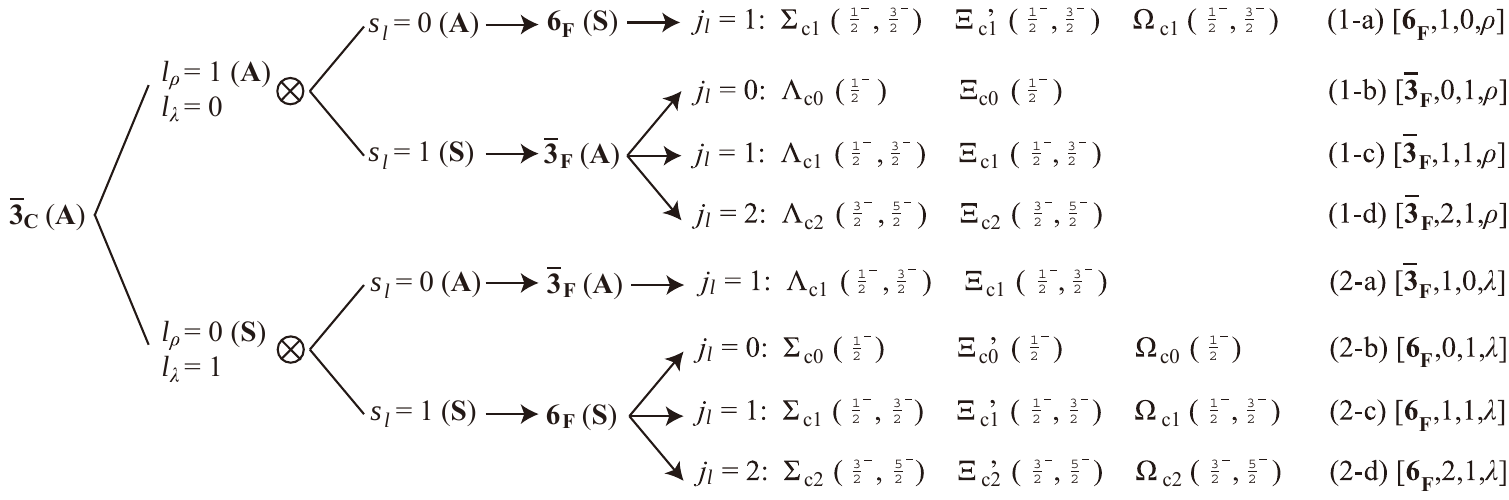}
\end{center}
\caption{Categorization of the $P$-wave single charmed baryons.}
\label{fig:pwave}
\end{figure*}

In this section we briefly introduce the notations and conventions used throughout this work. A singly charmed baryon is described as a three-quark system composed of two light $up/down/strange$ quarks and a heavy $charm$ quark. To properly understand the structure of singly charmed baryons, it is essential to analyze the quantum numbers and symmetries associated with the two light quarks, including their color, flavor, spin, and orbital degrees of freedom. These components are constrained by the Pauli exclusion principle, requiring the total wave function of the two light quarks to be antisymmetric under exchange. The structures are summarized as follows:
\begin{itemize}

\item The two light quarks must be in an antisymmetric color configuration, forming a color antitriplet ($\mathbf{\bar{3}}_C$).

\item The flavor wave function can be either antisymmetric, corresponding to the $SU(3)$ flavor antitriplet ($\mathbf{\bar{3}}_F$), or symmetric, corresponding to the $SU(3)$ flavor sextet ($\mathbf{6}_F$).

\item The spin configuration of the two light quarks can be antisymmetric ($s_l = 0$) or symmetric ($s_l = 1$), where $s_l$ denotes the total spin of the light diquark system.

\item The relative orbital excitation can also be either antisymmetric or symmetric. In the $\rho$-mode excitation ($l_\rho = 1$, $l_\lambda = 0$), the orbital angular momentum resides between the two light quarks and is antisymmetric. In the $\lambda$-mode excitation ($l_\rho = 0$, $l_\lambda = 1$), it is between the charm quark and the light diquark system and is symmetric. Here, $l_\rho$ and $l_\lambda$ denote the orbital angular momenta associated with the $\rho$- and $\lambda$-mode excitations, respectively.

\end{itemize}
As illustrated in Fig.~\ref{fig:pwave}, the $P$-wave charmed baryons can be systematically categorized into eight distinct HQET multiplets. Among them, four belong to the $SU(3)$ flavor antitriplet representation $\mathbf{\bar{3}}_F$, while the remaining four belong to the $SU(3)$ flavor sextet representation $\mathbf{6}_F$. We label each multiplet using the notation $[\mathrm{flavor}, j_l, s_l, \rho/\lambda]$, where $s_l$ denotes the total spin of the light diquark system, and $j_l = l\rho \otimes l\lambda \otimes s_l$ represents the total angular momentum of the light degrees of freedom. For each HQET multiplet, the total spin of the baryon is obtained by coupling $j_l$ with the heavy quark spin $s_Q = 1/2$, leading to physical states with $J = |j_l \pm 1/2|$.

In Refs.~\cite{Chen:2017sci,Yang:2021lce} we conducted a detailed investigation of the mass spectra and strong decay properties of $P$-wave charmed baryons using QCD sum rules and light-cone sum rules within the HQET framework. In the present work we extend this analysis to explore their radiative decay properties, employing light-cone sum rules still within the HQET framework. The mass spectra obtained in our previous studies will be used as input parameters for the current analysis, and are summarized in Table~\ref{tabmass}.

\begin{table*}[ht]
\renewcommand{\arraystretch}{1.6}
\caption{Parameters of $P$-wave singly charmed baryons in the $SU(3)$ flavor $\bf6_F$ representation. Decay constants in the last column satisfy $f_{\Sigma^{++}_c} = f_{\Sigma^0_c} = \sqrt2 f_{\Sigma^+_c}$ and $f_{\Xi^{\prime+}_c} = f_{\Xi^{\prime0}_c}$. Taken from Refs.~\cite{Chen:2017sci,Yang:2021lce}.}
\label{tabmass}
\begin{tabular}{ c |c | c | c | c | c c | c | c}
\hline\hline
\multirow{2}{*}{doublets}&\multirow{2}{*}{~~B~~} & $\omega_c$ & ~~~Working region~~~ & ~~~~~~~$\overline{\Lambda}$~~~~~~~ & ~~~Baryon~~~ & ~~~~Mass~~~~~ & ~Difference~ & Decay constant
\\    &                                           & (GeV) & (GeV)      & (GeV)                &                               ($j^P$)       & (GeV)      & (MeV)        & (GeV$^{4}$)
\\ \hline\hline
\multirow{6}{*}{$[\mathbf{6}_F, 1, 0, \rho]$}
& \multirow{2}{*}{$\Sigma_c$} & \multirow{2}{*}{$1.74\pm0.10$} & \multirow{2}{*}{$0.27< T < 0.32$} & \multirow{2}{*}{$1.25 \pm 0.11$} & $\Sigma_c(1/2^-)$ & $2.77 \pm 0.14$ & \multirow{2}{*}{$15 \pm 6$} & $0.067 \pm 0.017~(\Sigma^+_c(1/2^-))$
\\ \cline{6-7}\cline{9-9}
& & & & & $\Sigma_c(3/2^-)$ & $2.79 \pm 0.14$ & &$0.031 \pm 0.008~(\Sigma^+_c(3/2^-))$
\\ \cline{2-9}
& \multirow{2}{*}{$\Xi^\prime_c$} & \multirow{2}{*}{$1.87\pm0.10$} & \multirow{2}{*}{$0.26< T < 0.34$} & \multirow{2}{*}{$1.36 \pm 0.10$} & $\Xi^\prime_c(1/2^-)$ & $2.88 \pm 0.14$ & \multirow{2}{*}{$13 \pm 5$} & $0.059 \pm 0.014~(\Xi^{\prime0}_c(1/2^-))$
\\ \cline{6-7}\cline{9-9}
& & & & & $\Xi^\prime_c(3/2^-)$ & $2.89 \pm 0.14$ & &$0.028 \pm 0.007~(\Xi^{\prime0}_c(3/2^-))$
\\ \cline{2-9}
& \multirow{2}{*}{$\Omega_c$} & \multirow{2}{*}{$2.00\pm0.10$} & \multirow{2}{*}{$0.26< T < 0.35$} & \multirow{2}{*}{$1.48 \pm 0.09$} & $\Omega_c(1/2^-)$ & $2.99 \pm 0.15$ & \multirow{2}{*}{$12 \pm 5$} & $0.105 \pm 0.023~(\Omega^0_c(1/2^-))$
\\ \cline{6-7}\cline{9-9}
& & & & & $\Omega_c(3/2^-)$ & $3.00 \pm 0.15$ & &$0.049 \pm 0.011~(\Omega^0_c(3/2^-))$
\\ \hline
\multirow{3}{*}{$[\mathbf{6}_F, 0, 1, \lambda]$} & $\Sigma_c$ & $1.35\pm0.10$ & $T=0.27$ & $1.10 \pm 0.04$ & $\Sigma_c(1/2^-)$ & $2.83 \pm 0.05$ & -- & $0.045 \pm 0.008~(\Sigma^+_c(1/2^-))$
\\ \cline{2-9}
                                                 & $\Xi^\prime_c$ & $1.57\pm0.10$ & $0.27< T < 0.29$ & $1.22 \pm 0.08$ & $\Xi^\prime_c(1/2^-)$ & $2.90 \pm 0.13$ & -- & $0.041 \pm 0.009~(\Xi^{\prime0}_c(1/2^-))$
\\ \cline{2-9}
                                                 & $\Omega_c$ & $1.78\pm0.10$& $0.27< T < 0.31$ & $1.37 \pm 0.09$ & $\Omega_c(1/2^-)$ & $3.03 \pm 0.18$ & -- & $0.081 \pm 0.020~(\Omega^0_c(1/2^-))$
\\ \hline
\multirow{6}{*}{$[\mathbf{6}_F, 1, 1, \lambda]$}
& \multirow{2}{*}{$\Sigma_c$} & \multirow{2}{*}{$1.72\pm0.10$} & \multirow{2}{*}{$T=0.33$} & \multirow{2}{*}{$1.03 \pm 0.12$} & $\Sigma_c(1/2^-)$ & $2.73 \pm 0.17$ & \multirow{2}{*}{$41 \pm 16$} & $0.045 \pm 0.011~(\Sigma^+_c(1/2^-))$
\\ \cline{6-7}\cline{9-9}
& & & & & $\Sigma_c(3/2^-)$ & $2.77 \pm 0.17$ & &$0.021 \pm 0.005~(\Sigma^+_c(3/2^-))$
\\ \cline{2-9}
& \multirow{2}{*}{$\Xi^\prime_c$} & \multirow{2}{*}{$1.72\pm0.10$} & \multirow{2}{*}{$T=0.34$} & \multirow{2}{*}{$1.14 \pm 0.09$} & $\Xi^\prime_c(1/2^-)$ & $2.91 \pm 0.12$ & \multirow{2}{*}{$38 \pm 14$} & $0.041 \pm 0.008~(\Xi^{\prime0}_c(1/2^-))$
\\ \cline{6-7}\cline{9-9}
& & & & & $\Xi^\prime_c(3/2^-)$ & $2.95 \pm 0.12$ & &$0.019 \pm 0.004~(\Xi^{\prime0}_c(3/2^-))$
\\ \cline{2-9}
& \multirow{2}{*}{$\Omega_c$} & \multirow{2}{*}{$1.72\pm0.10$} & \multirow{2}{*}{$T=0.35$} & \multirow{2}{*}{$1.22 \pm 0.07$} & $\Omega_c(1/2^-)$ & $3.04 \pm 0.10$ & \multirow{2}{*}{$36\pm 13$} & $0.069 \pm 0.011~(\Omega^0_c(1/2^-))$
\\ \cline{6-7}\cline{9-9}
& & & & & $\Omega_c(3/2^-)$ & $3.07 \pm 0.09$ & &$0.032 \pm 0.005~(\Omega^0_c(3/2^-))$
\\ \hline
\multirow{6}{*}{$[\mathbf{6}_F, 2, 1, \lambda]$}
& \multirow{2}{*}{$\Sigma_c$} & \multirow{2}{*}{$1.50\pm0.10$} & \multirow{2}{*}{$0.28< T < 0.29$} & \multirow{2}{*}{$1.09 \pm 0.09$} & $\Sigma_c(3/2^-)$ & $2.78 \pm 0.13$ & \multirow{2}{*}{$86 \pm 36$} & $0.055 \pm 0.013~(\Sigma^+_c(3/2^-))$
\\ \cline{6-7}\cline{9-9}
& & & & & $\Sigma_c(5/2^-)$ & $2.87 \pm 0.11$ & &$0.033 \pm 0.008 ~(\Sigma^+_c(5/2^-))$
\\ \cline{2-9}
 & \multirow{2}{*}{$\Xi^\prime_c$} & \multirow{2}{*}{$1.72\pm0.10$} & \multirow{2}{*}{$0.27< T < 0.32$} & \multirow{2}{*}{$1.24 \pm 0.12$} & $\Xi^\prime_c(3/2^-)$ & $2.96 \pm 0.20$ & \multirow{2}{*}{$66 \pm 27$} & $0.057 \pm0.016 ~(\Xi^{\prime0}_c(3/2^-))$
\\ \cline{6-7}\cline{9-9}
& & & & & $\Xi^\prime_c(5/2^-)$ & $3.02 \pm 0.18$ & &$0.034 \pm 0.009 ~(\Xi^{\prime0}_c(5/2^-))$
\\ \cline{2-9}
& \multirow{2}{*}{$\Omega_c$} & \multirow{2}{*}{$1.85\pm0.10$} & \multirow{2}{*}{$0.26< T < 0.33$} & \multirow{2}{*}{$1.35 \pm 0.11$} & $\Omega_c(3/2^-)$ & $3.08\pm 0.19$ & \multirow{2}{*}{$59 \pm 24$} & $0.103 \pm 0.026~(\Omega^0_c(3/2^-))$
\\ \cline{6-7}\cline{9-9}
& & & & & $\Omega_c(5/2^-)$ & $3.14 \pm 0.18$ & &$0.062 \pm 0.016~(\Omega^0_c(5/2^-))$
\\ \hline\hline
\end{tabular}
\end{table*}

\section{Radiative Decay Properties}
\label{sec:6FD}

In this section we investigate the radiative decay properties of $P$-wave charmed baryons belonging to the $SU(3)$ flavor $\mathbf{6}_F$ representation. The possible radiative decay channels under consideration include:
\begin{align}
&\Sigma_c^{+}\to\Lambda_c^+(\Sigma_c^{+},\Sigma_c^{*+})\gamma\, ,\\
&\Sigma_c^{++}\to\Sigma_c^{++}(\Sigma_c^{*++})\gamma\, ,\\
&\Sigma_c^{0}\to\Sigma_c^{0}(\Sigma_c^{*0})\gamma\, ,\\
&\Xi_c^{\prime+}\to\Xi_c^+(\Xi_c^{\prime+},\Xi_c^{*+})\gamma\, ,\\
&\Xi_c^{\prime0}\to\Xi_c^0(\Xi_c^{\prime0},\Xi_c^{*0})\gamma\, ,\\
&\Omega_c^{0}\to\Omega_c^0(\Omega_c^{*0})\gamma\, .
\end{align}
The corresponding transition amplitudes are given by~\cite{Ivanov:1999bk}:
\begin{align}\label{amp1}
&\mathcal{M}(X_c({1/2}^-)\to Y_c({1/2}^+)\gamma)
\\ \nonumber
&~~~~~~~~~~~~~~~~~
=\frac{1}{\sqrt 3}g\bar X_c[g^{\mu\nu} v \cdot q-v^\mu q^\nu]\gamma_\nu\gamma_5 Y_c\epsilon^*_\mu \, ,
\\
&\mathcal{M}(X_c({1/2}^-)\to Y_c({3/2}^+)\gamma)
\\ \nonumber
&~~~~~~~~~~~~~~~~~
=g\bar X_c[g^{\mu\nu} v \cdot q-v^\mu q^\nu]Y_c^{\nu}\epsilon^*_\mu \, ,
\\
&\mathcal{M}(X_c({3/2}^-)\to Y_c({1/2}^+)\gamma)
\\ \nonumber
&~~~~~~~~~~~~~~~~~
=g\bar X_c^{\nu}[g^{\mu\nu} v \cdot q-v^\mu q^\nu]Y_c\epsilon^*_\mu \, ,
\\
&\mathcal{M}(X_c({3/2}^-)\to Y_c({3/2}^+)\gamma)
\\ \nonumber
&~~~~~~~~~~~~~~~~~
=\frac{1}{\sqrt 3}g\bar X_c^\alpha[g^{\mu\nu} v \cdot q-v^\mu q^\nu]\gamma_\nu\gamma_5Y_{c\alpha}\epsilon^*_\mu \, ,
\\
&\mathcal{M}(X_c({5/2}^-)\to Y_c({3/2}^+)\gamma)
\\ \nonumber
&~~~~~~~~~~~~~~~~~
=g\bar X_c^{\alpha\nu}[g^{\mu\nu} v \cdot q-v^\mu q^\nu]Y_{c\alpha}\epsilon^*_\mu \, ,
\end{align}
where $X_c^{(\mu\nu)}$ denotes the initial $P$-wave charmed baryon, $Y_c^{(\mu)}$ denotes the final ground-state charmed baryon, and $\epsilon_\mu$ is the polarization vector of the emitted photon.

The radiative decay width can be calculated using the following formula:
\begin{equation}
\Gamma(X_c\to Y_c\gamma)=\frac{1}{2J+1}\frac{|\vec q|}{8\pi M_{X_b}^2}\sum\limits_{spins} {|\mathcal{M}(X_c\to Y_c\gamma)|^2}\, ,
\end{equation}
where $M_{X_c}$ denotes the mass of the initial charmed baryon, and $\vec q$ represents the three-momentum of the final charmed baryon in the rest frame of the decaying particle.

As an illustrative example, we begin by analyzing the radiative decay of the $\Xi_c^{\prime+}({1/2}^-)$, which belongs to the $[\mathbf{6}_F, 1, 1, \lambda]$ doublet, into $\Xi_c^{*+}(3/2^+)$ accompanied by a photon. To study this transition, we consider the following two-point correlation function:
\begin{align} \nonumber
\Pi^\mu(\omega, \, \omega^\prime) &=\int d^4 x e^{-i k \cdot x} \langle 0 | J_{\Xi_c^{\prime+}[{\frac{1}{2}^-}],1,1,\lambda}(0) \bar J^\mu_{\Xi_c^{*+}}(x) | \gamma \rangle
\\
&={1+v\!\!\!\slash\over2} \epsilon^*_\mu G_{\Xi_c^{\prime+}[{1\over2}^-] \rightarrow \Xi_c^{*+}\gamma} (\omega, \omega^\prime) \, ,
\end{align}
where
\begin{eqnarray}
\label{ww}
k^\prime = k + q \, ,  \, \omega^\prime = v \cdot k^\prime \, ,  \, \omega = v \cdot k \, ,
\end{eqnarray}
with $k^\prime_\mu$, $k_\mu$, and $q_\mu$ denoting the four-momenta of the initial baryon, final baryon, and photon, respectively.

\begin{widetext}
The interpolating fields $ J_{\Xi_c^{\prime+}[{\frac{1}{2}^-}],1,1,\lambda}(x)$ and $J^\mu_{\Xi_c^{*+}}(x)$ have been systematically constructed in Refs.~\cite{Mao:2015gya,Liu:2007fg} as follows:
\begin{align}
J_{\Xi_c^{\prime+}[{\frac{1}{2}^-}],1,1,\lambda}(x)
&= i \epsilon_{abc} \Big ( [\mathcal{D}_t^{\mu} u^{aT}(x)] C \gamma_t^\nu s^b(x) +  u^{aT}(x) C \gamma_t^\nu [\mathcal{D}_t^{\mu} s^b(x)] \Big ) \sigma_t^{\mu\nu} h_v^c(x) \, ,
\\
J^\mu_{\Xi_c^{*+}}(x)
&=\epsilon_{abc}[u^{aT}(x)C\gamma_{\nu}s^{b}(x)](-g^{\mu\nu}_t+\frac{1}{3}\gamma^\mu_t\gamma^\nu_t)h_{v}^{c}(x)\, ,
\end{align}
where \( h_v^c(x) \) denotes the heavy quark field.

At the hadronic level, the function $G_{\Xi_c^{\prime+}[{1\over2}^-] \rightarrow \Xi_c^{*+}\gamma}(\omega, \omega^\prime)$ contains the following pole contribution derived from the double dispersion relation:
\begin{align}
G_{\Xi_c^{\prime+}[{1\over2}^-] \rightarrow \Xi_c^{*+}\gamma} (\omega, \omega^\prime)
= g_{\Xi_c^{\prime+}[{1\over2}^-] \rightarrow \Xi_c^{*+}\gamma} \times { f_{\Xi_c^{\prime+}[{1\over2}^-]} f_{\Xi_c^{*+}} \over (\bar \Lambda_{\Xi_c^{\prime+}[{1\over2}^-]} - \omega^\prime) (\bar \Lambda_{\Xi_c^{*+}} - \omega)}+\cdot\cdot\cdot \label{G0C}\, ,
\end{align}
where the ellipsis denotes contributions from higher excited states and continuum.

At the quark-gluon level, the two-point correlation function $\Pi^\mu(\omega, \omega^\prime)$ can be evaluated using the operator product expansion (OPE). Through this procedure, we extract the invariant function $G_{\Xi_c^{\prime+}[{1\over2}^-] \rightarrow \Xi_c^{*+}\gamma}(\omega, \omega^\prime)$, which encapsulates the short-distance dynamics of the transition:
\begin{align}
\label{Gt1}
&G_{\Xi_c^{\prime+}[{1\over2}^-] \rightarrow \Xi_c^{*+}\gamma} (\omega, \omega^\prime)
\\ \nonumber
&=\int_0^\infty  {dt} \int_0^1 {du{e^{i(1 - u)\omega 't}}{e^{iu\omega t}}}\times 8 \Bigg(\frac{{{e_s}{f_{3\gamma }}{\psi ^\alpha }({u})u v\cdot q}}{{48{\pi ^2}{t^2}}}-\frac{{{e_s}{\phi _\gamma }({u})\chi uv\cdot q}}{{72}}\left\langle {\bar qq}\right\rangle \left\langle {\bar ss} \right\rangle-\frac{{{e_s}{\phi_\gamma }({u})\chi t^2 u v\cdot q}}{{1152}}\left\langle {g_s\bar q\sigma Gs} \right\rangle\left\langle {\bar ss} \right\rangle
\\\nonumber
&+\frac{{{e_u}{f_{3\gamma }}{\psi ^\alpha }({u})uv\cdot q}}{{48{\pi ^2}{t^2}}}- \frac{{{e_u}{\phi_\gamma }({u})\chi {m_s}uv\cdot q}}{{24{\pi ^2}{t^2}}}\left\langle {\bar qq} \right\rangle-\frac{{{e_u}{\phi _\gamma }({u})\chi uv\cdot q}}{{72}}\left\langle {\bar qq} \right\rangle \left\langle {ss} \right\rangle
+\frac{{{e_u}{f_{3\gamma }}{\psi ^\alpha }({u}){m_s}t^2 u v\cdot q}}{{1152}}\left\langle {\bar ss} \right\rangle
\\\nonumber
&-\frac{{{e_u}{\phi _\gamma }({u})\chi t^2uv\cdot q}}{{1152}}\left\langle {\bar qq} \right\rangle \left\langle {g_s\bar s\sigma Gs} \right\rangle\Bigg)
-\int_0^\infty {dt} \int_0^1 {du\mathcal{D}{\underline\alpha}{e^{i\omega't({\alpha_2} + u{\alpha_3})}}{e^{i\omega t(1 - {\alpha_2} - u{\alpha_3})}}}\Bigg(-
\frac{{{f_{3\gamma }}\mathcal A(\underline\alpha)wv\cdot q}}{{24{\pi ^2}{t^2}}}+ \frac{{{f_{3\gamma }}\mathcal V(\underline\alpha)v\cdot q}}{{24{\pi ^2}{t^2}}}
\\\nonumber
&- \frac{{{f_{3\gamma }}\mathcal A(\underline\alpha)v\cdot q}}{{24{\pi^2}{t^2}}} - \frac{{{f_{3\gamma }}\mathcal V(\underline\alpha)wv\cdot q}}{{8{\pi^2}{t^2}}}
- \frac{{i{f_{3\gamma }}\mathcal A(\underline\alpha)(v\cdot q)^2}}{{24{\pi ^2}{t}}} - \frac{{i{f_{3\gamma }}\mathcal V(\underline\alpha)(v\cdot q)^2}}{{24{\pi ^2}{t}}} + \frac{{i{f_{3\gamma }}\mathcal A(\underline\alpha)w(v\cdot q)^2}}{{24{\pi ^2}{t}}} - \frac{{i{f_{3\gamma }}\mathcal V(\underline\alpha)w(v\cdot q)^2}}{{24{\pi ^2}{t}}}
\\\nonumber
&- \frac{{i{f_{3\gamma }}\mathcal {A}(\underline\alpha){\alpha_2}(v\cdot q)^2}}{{24{\pi ^2}{t}}} - \frac{{i{f_{3\gamma }}\mathcal V(\underline\alpha){\alpha _2}(v\cdot q)^2}}{{24{\pi ^2}{t}}} - \frac{{i{f_{3\gamma }}\mathcal{A}(\underline{\alpha})w{\alpha _2}(v\cdot q)^2}}{{24{\pi ^2}{t}}} - \frac{{i{f_{3\gamma }}\mathcal{V}(\underline{\alpha})w{\alpha _2}(v\cdot q)^2}}{{24{\pi ^2}{t}}}\\ \nonumber
&- \frac{{i{f_{3\gamma }}\mathcal \mathcal{A}(\underline\alpha)w{\alpha _3}(v\cdot q)^2}}{{24{\pi ^2}{t}}} - \frac{{i{f_{3\gamma }}\mathcal V(\underline\alpha)w{\alpha _3}(v\cdot q)^2}}{{24{\pi ^2}{t}}} - \frac{{i{f_{3\gamma }}\mathcal{A}(\underline\alpha){w^2}{\alpha _3}(v\cdot q)^2}}{{24{\pi ^2}{t}}} - \frac{{i{f_{3\gamma }}\mathcal V(\underline\alpha){w^2}{\alpha _3}(v\cdot q)^2}}{{24{\pi ^2}{t}}}\Bigg)(e_s-e_u)\, .
\end{align}

After performing the double Borel transformation with respect to the variables $\omega$ and $\omega^\prime$, introducing the Borel parameters $T_1$ and $T_2$, respectively, we obtain the following expression for the transformed correlation function:
\begin{align}
\label{Gt2}
& g_{\Xi_c^{\prime+}[{1\over2}^-] \rightarrow \Xi_c^{*+}\gamma} \times{ f_{\Xi_c^+[{1\over2}^-]} f_{\Xi_c^{*+}} \over(\bar \Lambda_{\Xi_c^+[{1\over2}^-]}-\omega^\prime)(\bar\Lambda_{\Xi_c^{*+}}-\omega)}
\\ \nonumber
&= 8\times\Bigg(-\frac{{{e_s}\chi }}{{72}}\left\langle {\bar qq} \right\rangle \left\langle {\bar ss} \right\rangle {(iT)^2}{f_1}(\frac{{{\omega _c}}}{T})\frac{\partial }{{\partial {u_0}}}{\phi _\gamma }({u_0}){u_0}-\frac{{{e_s}\chi }}{{1152}}\left\langle {g_s\bar q\sigma Gq} \right\rangle \left\langle {\bar ss} \right\rangle \frac{\partial }{{\partial {u_0}}}{\phi _\gamma }({u_0}){u_0}
- \frac{{{e_u}{f_{3\gamma }}{m_s}}}{{1152}}\left\langle {\bar ss} \right\rangle \frac{\partial }{{\partial {u_0}}}{\psi ^\alpha }({u_0}){u_0}
\\\nonumber
&- \frac{{{e_u}\chi }}{{1152}}\left\langle {\bar qq} \right\rangle \left\langle {g_s\bar s\sigma Gs} \right\rangle \frac{\partial }{{\partial {u_0}}}{\phi _\gamma }({u_0}){u_0}-\frac{{{e_s}{f_{3\gamma }}}}{{48{\pi ^2}}}{(iT)^4}{f_3}(\frac{{{\omega _c}}}{T})\frac{\partial }{{\partial {u_0}}}{\psi ^\alpha }({u_0}){u_0}
+\frac{{{e_u}{f_{3\gamma }}}}{{48{\pi ^2}}}{(iT)^4}{f_3}(\frac{{{\omega _c}}}{T})\frac{\partial }{{\partial {u_0}}}{\psi ^\alpha }({u_0}){u_0}
\\\nonumber
&- \frac{{{e_u}\chi }}{{72}}\left\langle {\bar qq} \right\rangle \left\langle {\bar ss} \right\rangle {(iT)^2}{f_1}(\frac{{{\omega _c}}}{T})\frac{\partial }{{\partial {u_0}}}{\phi _\gamma }({u_0}){u_0}
-\frac{{{e_u}\chi {m_s}}}{{24{\pi ^2}}}\left\langle {\bar qq} \right\rangle {(iT)^4}{f_3}(\frac{{{\omega _c}}}{T})\frac{\partial }{{\partial {u_0}}}{\phi _\gamma }({u_0}){u_0}\Bigg)
\\\nonumber
&-\Bigg(-\frac{{{f_{3\gamma }}}}{{8{\pi ^2}{u_0}}}{(iT)^4}{f_3}(\frac{{{\omega _c}}}{T})\int_0^{\frac{1}{2}}{d\alpha_2} \int_{\frac{1}{2}-\alpha_2}^{1-\alpha_2}{d\alpha_3}\Big(\frac{1}{{3{\alpha_3}}}\frac{\partial }{{\partial {a_3}}}\mathcal A(\underline\alpha) - \frac{1}{{3{a_3}}}\frac{\partial }{{\partial {a_3}}}\mathcal V(\underline\alpha)
+ \frac{{{u_0}}}{{3{a_3}}}\frac{\partial }{{\partial {\alpha_3}}}\mathcal A(\underline\alpha) + \frac{{{u_0}}}{{{a_3}}}\frac{\partial }{{\partial {a_3}}}\mathcal V(\underline\alpha)\Big)
\\\nonumber
&+\frac{{{f_{3\gamma }}}}{{24{\pi ^2}u_0^2}}{(iT)^4}{f_3}(\frac{{{\omega _c}}}{T})\int_0^{\frac{1}{2}}{d\alpha_2} \int_{\frac{1}{2}-\alpha_2}^{1-\alpha_2}{d\alpha_3}\Big(
  \frac{1}{{{\alpha_3}}}\frac{{{\partial ^2}}}{{\partial \alpha_3^2}}\mathcal A(\underline\alpha) - \frac{1}{{{\alpha_3}}}\frac{{{\partial ^2}}}{{\partial \alpha_3^2}}\mathcal V(\underline\alpha) + \frac{{{u_0}}}{{{\alpha_3}}}\frac{{{\partial ^2}}}{{\partial \alpha_3^2}}\mathcal A(\underline\alpha) - \frac{{i{u_0}}}{{{\alpha_3}}}\frac{{{\partial ^2}}}{{\partial \alpha_3^2}}\mathcal V(\underline\alpha)
\\\nonumber
&+ \frac{{{\alpha _2}}}{{{\alpha_3}}}\frac{{{\partial ^2}}}{{\partial \alpha_3^2}}\mathcal A(\underline\alpha) - \frac{{{\alpha _2}}}{{{\alpha_3}}}\frac{{{\partial ^2}}}{{\partial \alpha_3^2}}\mathcal V(\underline\alpha)
 - \frac{{{u_0}{\alpha _2}}}{{{\alpha_3}}}\frac{{{\partial ^2}}}{{\partial \alpha_3^2}}\mathcal A(\underline\alpha) + \frac{{{u_0}{\alpha _2}}}{{{\alpha_3}}}\frac{{{\partial ^2}}}{{\partial \alpha_3^2}}\mathcal V(\underline\alpha)
- \frac{{{u_0}{\alpha _3}}}{{{\alpha_3}}}\frac{{{\partial ^2}}}{{\partial \alpha_3^2}}\mathcal A(\underline\alpha) - \frac{{{u_0}{\alpha _3}}}{{{\alpha_3}}}\frac{{{\partial ^2}}}{{\partial \alpha_3^2}}\mathcal V(\underline\alpha)
\\\nonumber
&-\frac{{u_0^2{\alpha _3}}}{{{\alpha_3}}}\frac{{{\partial ^2}}}{{\partial \alpha_3^2}}\mathcal A(\underline\alpha) - \frac{{u_0^2{\alpha _3}}}{{{\alpha_3}}}\frac{{{\partial ^2}}}{{\partial \alpha_3^2}}\mathcal V(\underline\alpha)\Big)\Bigg)(e_s-e_u)\, .
\end{align}
In the above expressions, $u_0 = {T_1 \over T_1 + T_2}$, $T = {T_1 T_2 \over T_1 + T_2}$, $f_n(x) = 1 - e^{-x} \sum_{k=0}^n {x^k \over k!}$, and $e_{u/d/s}$ denotes the electric charge of the up/down/strange quark, respectively.
\end{widetext}

\begin{figure*}[]
\subfigure[]{
\scalebox{0.6}{\includegraphics{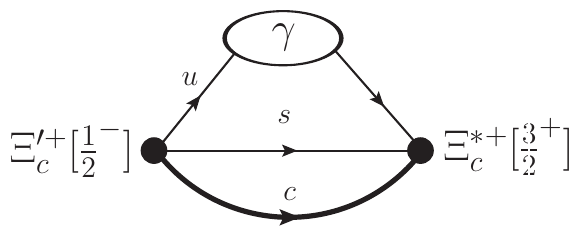}}}
\subfigure[]{
\scalebox{0.6}{\includegraphics{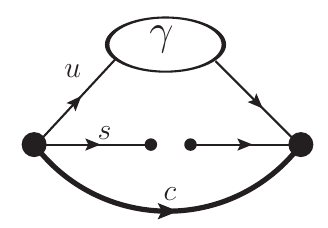}}}
\subfigure[]{
\scalebox{0.6}{\includegraphics{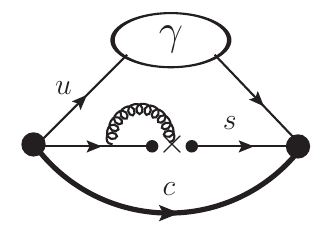}}}
\subfigure[]{
\scalebox{0.6}{\includegraphics{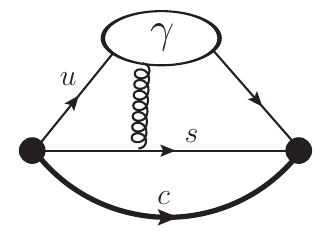}}}
\\
\subfigure[]{
\scalebox{0.6}{\includegraphics{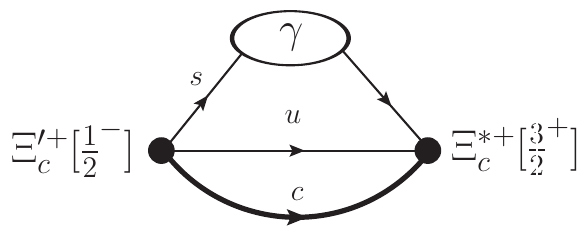}}}
\subfigure[]{
\scalebox{0.6}{\includegraphics{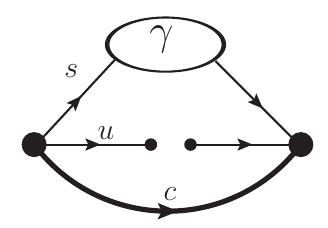}}}
\subfigure[]{
\scalebox{0.6}{\includegraphics{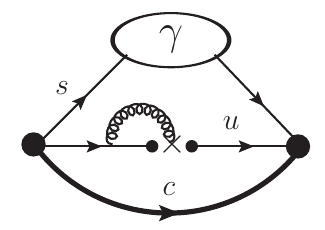}}}
\subfigure[]{
\scalebox{0.6}{\includegraphics{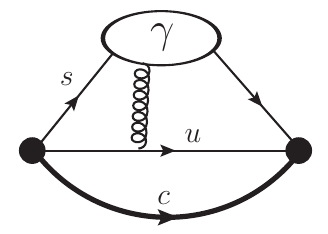}}}
\caption{Feynman diagrams for the process $\Xi_c^{\prime+}({1/2}^-)\to\Xi_c^{*+}(3/2^+)\gamma$.}
\label{fig:feynman}
\end{figure*}

\begin{figure}[hbt]
\begin{center}
\scalebox{0.95}{\includegraphics{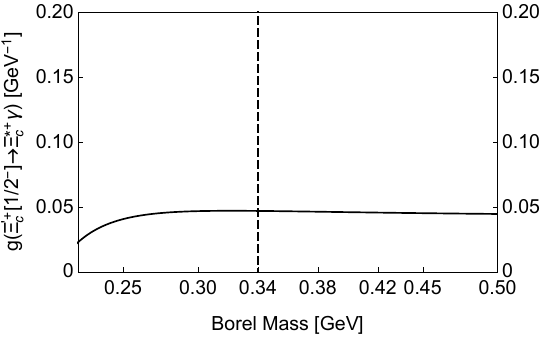}}
\caption{The coupling constant $g_{\Xi^{\prime+}_c[\frac{1}{2}^-]\to\Xi_c^{*+}\gamma}$ with respect to the Borel mass $T$. Here the charmed baryon $\Xi_c^{+}({1/2}^-)$ belonging to the $[\mathbf{6}_F, 1, 1, \lambda]$ doublet is investigated.
\label{fig:es}}
\end{center}
\end{figure}

Eq.~(\ref{Gt1}) and Eq.~(\ref{Gt2}) involve numerous light-cone distribution amplitudes, whose definitions and explicit expressions are systematically presented in Refs.~\cite{Ball:1998je,Ball:2006wn,Ball:2004rg,Ball:1998kk,Ball:1998sk,Ball:1998ff,Ball:2007rt,Ball:2007zt}. In our calculations, we include the Feynman diagrams shown in Fig.~\ref{fig:feynman}, encompassing the perturbative contributions as well as the nonperturbative effects from the quark condensate, the quark-gluon mixed condensate, and their combinations. Throughout this work, we adopt the renormalization scale of 1~GeV, at which the relevant QCD condensates take the following values~\cite{Yang:1993bp,Hwang:1994vp,Ovchinnikov:1988gk,Narison:2002woh,Jamin:2002ev,Ioffe:2002be,Shifman:2001ck,Gimenez:2005nt}:
\begin{eqnarray}
\nonumber  \langle \bar qq \rangle  &=& - (0.24 \mbox{ GeV})^3 \, ,
\\ \nonumber  \langle \bar ss \rangle &=& (0.8\pm 0.1)\times \langle\bar qq \rangle \, ,
\\ \langle g_s^2GG\rangle&=&(0.48\pm 0.14) \mbox{ GeV}^4\, ,
\label{eq:condensates}
\\ \nonumber  \langle g_s \bar q \sigma G q \rangle &=& M_0^2 \times \langle \bar qq \rangle\, ,
\\ \nonumber \langle g_s \bar s \sigma G s \rangle &=& M_0^2 \times \langle \bar ss \rangle\, ,
\\ \nonumber M_0^2 &=& 0.8 \mbox{ GeV}^2\, .
\end{eqnarray}
We perform the analysis at the symmetric point $T_1 = T_2 = 2T$, which corresponds to $u_0 =1/2$. Under this condition, the coupling constant $g_{\Xi_c^{\prime+}[{1\over2}^-] \to \Xi_c^{*+}\gamma}$ depends on two free parameters: the threshold value $\omega_c$ and the Borel mass $T$. Following the setup listed in Table~\ref{tabmass}, we choose the central value $T = 0.34$~GeV and $\omega_c = 1.72$ GeV, and plot the dependence of $g_{\Xi_c^{\prime+}[{1\over2}^-] \to \Xi_c^{*+}\gamma}$ on $T$ in Fig.~\ref{fig:es}. From this analysis, we obtain the coupling constant and corresponding decay width as
\begin{align}
g_{\Xi_c^{\prime+}[\frac{1}{2}^-]\to\Xi_c^{*+}[\frac{3}{2}^+]\gamma}&=0.047^{+0.016}_{-0.014} \, ,
\\
\Gamma_{\Xi_c^{\prime+}[\frac{1}{2}^-]\to\Xi_c^{*+}[\frac{3}{2}^+]\gamma}&=6.9^{+13,3}_{-~6.6}~\rm keV\, .
\end{align}
The quoted uncertainties originate from the variation of the threshold value $\omega_c$, the Borel mass $T$, the QCD vacuum condensates, and the input parameters of the light-cone distribution amplitudes used in the sum rules.

Following the procedures outlined above, we systematically investigate the radiative transitions of the four $SU(3)$ flavor $\mathbf{6}_F$ multiplets: $[\mathbf{6}_F, 1, 0, \rho]$, $[\mathbf{6}_F, 0, 1, \lambda]$, $[\mathbf{6}_F, 1, 1, \lambda]$, and $[\mathbf{6}_F, 2, 1, \lambda]$. Their members are categorized as follows:
\begin{itemize}

\item The $[\mathbf{6}_F, 1, 0, \rho]$ doublet contains
$\Sigma_c^+({1/2}^-)$, $\Sigma_c^+({3/2}^-)$, $\Sigma_c^{++}({1/2}^-)$, $\Sigma_c^{++}({3/2}^-)$, $\Sigma_c^0({1/2}^-)$, $\Sigma_c^0({3/2}^-)$, $\Xi_c^{\prime+}({1/2}^-)$,
$\Xi_c^{\prime+}({3/2}^-)$, $\Xi_c^{\prime0}({1/2}^-)$, $\Xi_c^{\prime0}({3/2}^-)$, $\Omega_c^0({1/2}^-)$, and $\Omega_c^0({3/2}^-)$.

\item The $[\mathbf{6}_F, 0, 1, \lambda]$ singlet contains
$\Sigma_c^+({1/2}^-)$, $\Sigma_c^{++}({1/2}^-)$,
$\Sigma_c^0({1/2}^-)$, $\Xi_c^{+}({1/2}^-)$,
$\Xi_c^{\prime0}({1/2}^-)$, and $\Omega_c^0({1/2}^-)$.

\item The $[\mathbf{6}_F, 1, 1, \lambda]$ doublet contains $\Sigma_c^+({1/2}^-)$, $\Sigma_c^+({3/2}^-)$, $\Sigma_c^{++}({1/2}^-)$, $\Sigma_c^{++}({3/2}^-)$, $\Sigma_c^0({1/2}^-)$, $\Sigma_c^0({3/2}^-)$, $\Xi_c^{\prime+}({1/2}^-)$,
$\Xi_c^{\prime+}({3/2}^-)$, $\Xi_c^{\prime0}({1/2}^-)$, $\Xi_c^{\prime0}({3/2}^-)$, $\Omega_c^0({1/2}^-)$, and $\Omega_c^0({3/2}^-)$.

\item The $[\mathbf{6}_F, 2, 1, \lambda]$ doublet contains $\Sigma_c^+({3/2}^-)$, $\Sigma_c^+({5/2}^-)$, $\Sigma_c^{++}({3/2}^-)$, $\Sigma_c^{++}({5/2}^-)$, $\Sigma_c^0({3/2}^-)$, $\Sigma_c^0({5/2}^-)$, $\Xi_c^{\prime+}({3/2}^-)$,
$\Xi_c^{\prime+}({5/2}^-)$, $\Xi_c^{\prime0}({3/2}^-)$, $\Xi_c^{\prime0}({5/2}^-)$, $\Omega_c^0({3/2}^-)$, and $\Omega_c^0({5/2}^-)$.

\end{itemize}
We study their radiative transitions into the ground-state charmed baryons. The corresponding light-cone QCD sum rule expressions are provided in the supplementary file OPE.nb. The numerical results for the radiative decays of the $\Sigma_c$, $\Xi_c^\prime$, and $\Omega_c$ baryons are summarized in Table~\ref{decayc6S}, Table~\ref{decayc6X}, and Table~\ref{decayc6O}, respectively.

\begin{table*}[ht]
\begin{center}
\renewcommand{\arraystretch}{1.5}
\caption{Radiative decay widths of the $P$-wave charmed baryons $\Sigma_c$ belonging to the $SU(3)$ flavor $\mathbf{6}_F$  representation.}
\setlength{\tabcolsep}{0.1mm}{
\begin{tabular}{c| c | c | c | c | c | c|c }
\hline\hline
\multirow{2}{*}{Multiplets}&Baryon & ~~~~~~Mass~~~~~~ & ~~~Difference~~~ & \multirow{2}{*}{~~~~~~Decay channels~~~~~~}  & ~~~~Coupling~~~~ & ~~~~Width~~~~ & ~~~~Total~~~~
\\ & ($j^P$) & ({GeV}) & ({MeV}) & &({$\rm GeV^{-1}$})&({keV})&({keV})
\\ \hline\hline
~~\multirow{6}{*}{$[\mathbf{6}_F, 0, 1, \lambda]$}~~ &
~~\multirow{6}{*}{$\Sigma_c({1\over2}^-)$}~~     &\multirow{6}{*}{$2.83\pm 0.05$}      &\multirow{6}{*}{-}
&$\Sigma_c^+({1\over2}^-)\to\Sigma_c^+\gamma$          &$0.30^{+0.09}_{-0.04}$   &$360^{+300}_{-150}$               &\multirow{2}{*}{$990^{+770}_{-460}$}
\\
&&&&$\Sigma_c^+({1\over2}^-)\to \Sigma_c^{*+}\gamma$     &$0.36^{+0.09}_{-0.06}$  &$630^{+470}_{-310}$                    &
\\ \cline{5-8}
&&&&$\Sigma_c^{++}({1\over2}^-)\to\Sigma_c^{++}\gamma$    &$1.20^{+0.37}_{-0.10}$    &$5800^{+3900}_{-2600}$                   &\multirow{2}{*}{$22000^{+18000}_{-10000}$}
\\
&&&&$\Sigma_c^{++}({1\over2}^-)\to \Sigma_c^{*++}\gamma$   &$1.40^{+0.49}_{-0.25}$   &$16000^{+14000}_{-~7800}$                       &
\\ \cline{5-8}
&&&&$\Sigma_c^0({1\over2}^-)\to\Sigma_c^0\gamma$       &$0.62^{+0.16}_{-0.11}$     &$1500^{+1000}_{-~760}$        &\multirow{2}{*}{$4000^{+2900}_{-2000}$}
\\
&&&&$\Sigma_c^0({1\over2}^-)\to \Sigma_c^{*0}\gamma$    &$0.72^{+0.18}_{-0.13}$    &$2500^{+1900}_{-1200}$                        &
\\ \hline
\multirow{12}{*}{$[\mathbf{6}_F, 1, 1, \lambda]$} &
\multirow{6}{*}{$\Sigma_c({1\over2}^-)$}     &\multirow{6}{*}{$2.73\pm 0.17$}      &\multirow{12}{*}{$6\pm 3$}
& $\Sigma_c^+({1\over2}^-)\to\Sigma_c^+\gamma$           &$0.039^{+0.017}_{-0.014}$ & $2.6^{+7.4}_{-2.6}$                         &\multirow{3}{*}{ $10000^{+18000}_{-~8100}$ }
\\
&&&&$\Sigma_c^+({1\over2}^-)\to \Sigma_c^{*+}\gamma$     &$0.022^{+0.011}_{-0.010}$  &$0.80^{+3.35}_{-0.80}$                        &
\\
&&&&$\Sigma_c^+({1\over2}^-)\to \Lambda_c^{+}\gamma$     &$1.3^{+0.7}_{-00.3}$  &$10000^{+18000}_{-~8100}$                        &
\\ \cline{5-8}
&&&&$\Sigma_c^{++}({1\over2}^-)\to\Sigma_c^{++}\gamma$     &$0.16^{+0.07}_{-0.06}$  &$14^{+43}_{-14}$                         &\multirow{2}{*}{$27^{+100}_{-~27}$}
\\
&&&&$\Sigma_c^{++}({1\over2}^-)\to \Sigma_c^{*++}\gamma$   &$0.091^{+0.040}_{-0.033}$  & $13^{+57}_{-13}$                     &
\\ \cline{5-8}
&&&&$\Sigma_c^0({1\over2}^-)\to\Sigma_c^0\gamma$        &$0.079^{+0.033}_{-0.029}$   &$11^{+30}_{-11}$                         &\multirow{2}{*}{$15^{+45}_{-15}$}
\\
&&&&$\Sigma_c^0({1\over2}^-)\to \Sigma_c^{*0}\gamma$   &$0.046^{+0.020}_{-0.017}$    &$3.5^{+15.0}_{-~3.5}$                         &
\\ \cline{2-3}\cline{5-8}
&\multirow{6}{*}{$\Sigma_b({3\over2}^-)$}     &\multirow{6}{*}{$2.77\pm 0.17$}      &
&$\Sigma_c^+({3\over2}^-)\to\Sigma_c^+\gamma$       &$0.028^{+0.013}_{-0.011}$    &$2.1^{+4.6}_{-2.1}$             &\multirow{3}{*}{$24000^{+34000}_{-20000}$ }
\\
&&&&$\Sigma_c^+({3\over2}^-)\to \Sigma_c^{*+}\gamma$    &$0.033^{+0.017}_{-0.011}$   &$0.81^{+2.70}_{-0.81}$                    &
\\
&&&&$\Sigma_c^+({3\over2}^-)\to \Sigma_c^{*+}\gamma$    &$1.8^{+0.6}_{-0.5}$   &$24000^{+34000}_{-20000}$                    &
\\ \cline{5-8}
&&&&$\Sigma_c^{++}({3\over2}^-)\to\Sigma_c^{++}\gamma$    &$0.11^{+0.06}_{-0.04}$   &$30^{+75}_{-30}$                         &\multirow{2}{*}{$45^{+120}_{-~45}$}
\\
&&&&$\Sigma_c^{++}({3\over2}^-)\to \Sigma_c^{*++}\gamma$   &$0.14^{+0.05}_{-0.06}$   & $15^{+47}_{-15}$                        &
\\ \cline{5-8}
&&&&$\Sigma_c^0({3\over2}^-)\to\Sigma_c^0\gamma$       &$0.056^{+0.025}_{-0.020}$    &$7.9^{+18.9}_{-~7.9}$                        &\multirow{2}{*}{$11^{+30}_{-11}$}
\\
&&&&$\Sigma_c^0({3\over2}^-)\to \Sigma_c^{*0}\gamma$    &$0.067^{+0.031}_{-0.023}$   & $3.3^{+11.4}_{-~3.3}$                    &
\\ \hline
\multirow{12}{*}{$[\mathbf{6}_F, 2, 1, \lambda]$} &
\multirow{6}{*}{$\Sigma_c({3\over2}^-)$}     &\multirow{6}{*}{$2.78\pm 0.13$}      &\multirow{12}{*}{$12\pm 5$}
&$\Sigma_c^+({3\over2}^-)\to\Sigma_c^+\gamma$           &$0.098^{+0.017}_{-0.026}$    &$26^{+39}_{-22}$               &\multirow{2}{*}{$31^{+51}_{-28}$}
\\
&&&&$\Sigma_c^+({3\over2}^-)\to \Sigma_c^{*+}\gamma$    &$0.083^{+0.017}_{-0.022}$    &$5.7^{+11.6}_{-~5.5}$                 &
\\ \cline{5-7}
&&&&$\Sigma_c^{++}({3\over2}^-)\to\Sigma_c^{++}\gamma$     &$0.39^{+0.09}_{-0.01}$   &$420^{+620}_{-370}$               &\multirow{2}{*}{$500^{+810}_{-460}$}
\\
&&&&$\Sigma_c^{++}({3\over2}^-)\to \Sigma_c^{*++}\gamma$    &$0.33^{+0.01}_{-0.01}$  & $90^{+190}_{-~87}$                        &
\\ \cline{5-8}
&&&&$\Sigma_c^0({3\over2}^-)\to\Sigma_c^0\gamma$        &$0.20^{+0.04}_{-0.06}$    &$110^{+170}_{-100}$                 &\multirow{2}{*}{$130^{+220}_{-120}$}
\\
&&&&$\Sigma_c^0({3\over2}^-)\to \Sigma_c^{*0}\gamma$      &$0.16^{+0.05}_{-0.04}$  & $21^{+45}_{-20}$                        &
\\ \cline{2-3}\cline{5-8}
&\multirow{3}{*}{$\Sigma_c({5\over2}^-)$}     &\multirow{3}{*}{$2.87\pm 0.11$}      &
&$\Sigma_c^+({5\over2}^-)\to \Sigma_c^{*+}\gamma$        &$0.034^{+0.07}_{-0.09}$   & $5.8^{+6.8}_{-4.6}$                     &$5.8^{+6.8}_{-4.6}$
\\ \cline{5-8}
&&&&$\Sigma_c^{++}({5\over2}^-)\to \Sigma_c^{*++}\gamma$    &$0.14^{+0.03}_{-0.05}$  & $98^{+110}_{-~85}$                     & $98^{+110}_{-~85}$
\\ \cline{5-8}
&&&&$\Sigma_c^0({5\over2}^-)\to \Sigma_c^{*0}\gamma$     &$0.04^{+0.01}_{-0.01}$   & $8.5^{+9.2}_{-6.8}$                     &$8.5^{+9.2}_{-6.8}$
\\ \hline
\multirow{12}{*}{$[\mathbf{6}_F, 1, 0, \rho]$} &
\multirow{6}{*}{$\Sigma_c({1\over2}^-)$}     &\multirow{6}{*}{$2.77\pm 0.14$}      &\multirow{12}{*}{$3\pm 1$}
&$\Sigma_c^+({1\over2}^-)\to\Sigma_c^+\gamma$         &$0.083^{+0.033}_{-0.026}$  &$17^{+33}_{-16}$         &\multirow{2}{*}{$23^{+38}_{-22}$}
\\
&&&&$\Sigma_c^+({1\over2}^-)\to \Sigma_c^{*+}\gamma$  &$0.048^{+0.017}_{-0.015}$ & $6.2^{+14.6}_{-~6.2}$                      &
\\  \cline{5-8}
&&&&$\Sigma_c^{++}({1\over2}^-)\to\Sigma_c^{++}\gamma$       &$0.33^{+0.15}_{-0.10}$   &$1100^{+2200}_{-1000}$         &\multirow{2}{*}{$1500^{+3200}_{-1400}$}
\\
&&&&$\Sigma_c^{++}({1\over2}^-)\to \Sigma_c^{*++}\gamma$  &$0.19^{+0.10}_{-0.06}$ & $390^{+1000}_{-~390}$                      &
\\  \cline{5-8}
&&&&$\Sigma_c^0({1\over2}^-)\to\Sigma_c^0\gamma$       &$0.17^{+0.06}_{-0.06}$   &$290^{+540}_{-290}$         &\multirow{2}{*}{$390^{+790}_{-390}$}
\\
&&&&$\Sigma_c^0({1\over2}^-)\to \Sigma_c^{*0}\gamma$  &$0.096^{+0.040}_{-0.029}$  & $99^{+250}_{-~99}$                      &
\\ \cline{2-3} \cline{5-8}
&\multirow{6}{*}{$\Sigma_b({3\over2}^-)$}&\multirow{6}{*}{$2.79\pm 0.14$} &
&$\Sigma_c^+({3\over2}^-)\to\Sigma_c^+\gamma$      & $0.059^{+0.026}_{-0.017}$   &  $10.0^{+19.6}_{-~9.1}$     &\multirow{2}{*}{$19^{+41}_{-18}$}
\\
&&&&$\Sigma_c^+({3\over2}^-)\to \Sigma_c^{*+}\gamma$ &$0.10^{+0.05}_{-0.02}$ &  $9.2^{+21.4}_{-~8.9}$                         &
\\  \cline{5-8}
&&&&$\Sigma_c^{++}({1\over2}^-)\to\Sigma_c^{++}\gamma$      &$0.24^{+0.10}_{-0.08}$    &$680^{+1700}_{-~660}$         &\multirow{2}{*}{$1300^{+3100}_{-1200}$}
\\
&&&&$\Sigma_c^{++}({1\over2}^-)\to \Sigma_c^{*++}\gamma$  &$0.40^{+0.17}_{-0.10}$ & $590^{+1400}_{-~570}$                      &
\\  \cline{5-8}
&&&&$\Sigma_c^0({1\over2}^-)\to\Sigma_c^0\gamma$     &$0.12^{+0.05}_{-0.04}$     &$170^{+310}_{-160}$         &\multirow{2}{*}{$320^{+660}_{-300}$}
\\
&&&&$\Sigma_c^0({1\over2}^-)\to \Sigma_c^{*0}\gamma$  &$0.20^{+0.09}_{-0.05}$ & $150^{+350}_{-140}$                      &
\\ \hline\hline
\end{tabular}}
\label{decayc6S}
\end{center}
\end{table*}

\begin{table*}[ht]
\begin{center}
\renewcommand{\arraystretch}{1.5}
\caption{Radiative decay widths of the $P$-wave charmed baryons $\Xi_c^\prime$ belonging to the $SU(3)$ flavor $\mathbf{6}_F$  representation.}
\setlength{\tabcolsep}{0.1mm}{
\begin{tabular}{c| c | c | c | c | c | c|c }
\hline\hline
\multirow{2}{*}{Multiplets}&Baryon & ~~~~~~Mass~~~~~~ & ~~~Difference~~~ & \multirow{2}{*}{~~~~~~Decay channels~~~~~~}  & ~~~~Coupling~~~~ & ~~~~Width~~~~ & ~~~~Total~~~~
\\ & ($j^P$) & ({GeV}) & ({MeV}) & &({GeV$^{-1}$})&({keV})&({keV})
\\ \hline\hline
~~\multirow{4}{*}{$[\mathbf{ 6}_F, 0, 1, \lambda]$}~~ &
~~\multirow{4}{*}{$\Xi_c^{\prime}({1\over2}^-)$}~~     &\multirow{4}{*}{$2.90\pm 0.13$}      &\multirow{4}{*}{-}
&$\Xi_c^{\prime+}({1\over2}^-)\to\Xi_c^{\prime+}\gamma$       &$0.49^{+0.17}_{-0.15}$     &$650^{+1100}_{-~590}$                &\multirow{2}{*}{$1500^{+3100}_{-1400}$}
\\
&&&&$\Xi_c^{\prime+}({1\over2}^-)\to \Xi_c^{*+}\gamma$         &$0.56^{+0.21}_{-0.15}$    &$880^{+2000}_{-~860}$
\\ \cline{5-8}
&&&&$\Xi_c^{\prime0}({1\over2}^-)\to\Xi_c^{\prime0}\gamma$     &$0.78^{+0.31}_{-0.22}$    &$1600^{+2800}_{-1500}$                          &\multirow{2}{*}{$2400^{+4500}_{-2300}$}
\\
&&&&$\Xi_c^{\prime0}({1\over2}^-)\to \Xi_c^{*0}\gamma$         &$0.52^{+0.20}_{-0.18}$    &$760^{+1700}_{-~760}$
\\ \hline
~~\multirow{8}{*}{$[\mathbf{6}_F, 1, 1, \lambda]$} ~~&
~~\multirow{4}{*}{$\Xi_c^{\prime}({1\over2}^-)$}~~     &\multirow{4}{*}{$2.91\pm 0.12$}      &\multirow{8}{*}{$7\pm 2$}
&$\Xi_c^{\prime+}({1\over2}^-)\to\Xi_c^{{\prime}+}\gamma$    &$0.081^{+0.028}_{-0.023}$   &$19^{+26}_{-14}$                  &\multirow{3}{*}{$27000^{+29000}_{-20000}$  }
\\
&&&&$\Xi_c^{\prime+}({1\over2}^-)\to \Xi_c^{*+}\gamma$       &$0.047^{+0.016}_{-0.014}$      &$6.9^{+13.3}_{-~6.6}$                        &
\\
&&&&$\Xi_c^{\prime+}({1\over2}^-)\to \Xi_c^{+}\gamma$       &$2.1^{+0.6}_{-0.5}$      &$27000^{+29000}_{-20000}$                        &
\\  \cline{5-8}
&&&&$\Xi_c^{\prime0}({1\over2}^-)\to\Xi_c^{{\prime}0}\gamma$     &$0.11^{+0.05}_{-0.03}$  &$35^{+61}_{-30}$            &\multirow{3}{*}{$110^{+240}_{-100}$}
\\
&&&&$\Xi_c^{\prime0}({1\over2}^-)\to \Xi_c^{*0}\gamma$        &$0.065^{+0.026}_{-0.022}$     &$13^{+27}_{-12}$                        &
\\
&&&&$\Xi_c^{\prime0}({1\over2}^-)\to \Xi_c^{0}\gamma$        &$0.10^{+0.08}_{-0.08}$     &$61^{+150}_{-~61}$                        &
\\ \cline{2-3}\cline{5-8}
&~~\multirow{4}{*}{$\Xi_c^{\prime}({3\over2}^-)$}~~     &\multirow{4}{*}{$2.95\pm 0.12$}
&&$\Xi_c^{\prime+}({3\over2}^-)\to\Xi_c^{{\prime}+}\gamma$    &$0.057^{+0.016}_{-0.016}$   &$12.0^{+39.9}_{-~9.2}$                &\multirow{3}{*}{$55000^{+62000}_{-37000}$}
\\
&&&&$\Xi_c^{\prime+}({3\over2}^-)\to \Xi_c^{*+}\gamma$         &$0.068^{+0.026}_{-0.018}$    & $5.9^{+9.7}_{-5.2}$                       &
\\
&&&&$\Xi_c^{\prime+}({3\over2}^-)\to \Xi_c^{+}\gamma$         &$2.7^{+0.9}_{-0.6}$    & $55000^{+62000}_{-37000}$                       &
\\  \cline{5-8}
&&&&$\Xi_c^{\prime0}({3\over2}^-)\to\Xi_c^{{\prime}0}\gamma$    &$0.080^{+0.032}_{-0.028}$   & $25^{+37}_{-21}$                 &\multirow{3}{*}{$40^{+62}_{-34}$}
\\
&&&&$\Xi_c^{\prime0}({3\over2}^-)\to \Xi_c^{*0}\gamma$        &$0.097^{+0.043}_{-0.033}$     & $12^{+22}_{-11}$         &
\\
&&&&$\Xi_c^{\prime0}({3\over2}^-)\to \Xi_c^{0}\gamma$        &$0.02^{+0.01}_{-0.01}$     & $3.1^{+3.4}_{-2.0}$         &
\\ \hline
~~\multirow{6}{*}{$[\mathbf{ 6}_F, 2, 1, \lambda]$}~~
&~~\multirow{4}{*}{$\Xi_c^{\prime}({3\over2}^-)$}~~    &\multirow{4}{*}{$2.78\pm 0.13$}      &\multirow{8}{*}{$11\pm 5$}
&$\Xi_c^{\prime+}({3\over2}^-)\to\Xi_c^{{\prime}+}\gamma$      &$0.13^{+0.07}_{-0.06}$    &$72^{+180}_{-~72}$                         &\multirow{2}{*}{$92^{+240}_{-~92}$}
\\
&&&&$\Xi_c^{\prime+}({3\over2}^-)\to \Xi_c^{*+}\gamma$         &$0.12^{+0.04}_{-0.06}$    &$20^{+57}_{-20}$                        &
\\  \cline{5-8}
&&&&$\Xi_c^{\prime0}({3\over2}^-)\to\Xi_c^{{\prime}0}\gamma$   &$0.21^{+0.14}_{-0.11}$    & $190^{+510}_{-190}$                         &\multirow{2}{*}{$240^{+670}_{-240}$}
\\
&&&&$\Xi_c^{\prime0}({3\over2}^-)\to \Xi_c^{*0}\gamma$        &$0.19^{+0.10}_{-0.07}$     &$50^{+160}_{-~50}$                       &
\\ \cline{2-3}\cline{5-8}
&~~\multirow{2}{*}{$\Xi_c^{\prime}({5\over2}^-)$}~~    &\multirow{2}{*}{$2.87\pm 0.11$}
&&$\Xi_c^{\prime+}({5\over2}^-)\to \Xi_c^{*+}\gamma$       &$0.056^{+0.026}_{-0.022}$      & $19^{+41}_{-19}$                      &$19^{+41}_{-19}$
\\  \cline{5-8}
&&&&$\Xi_c^{\prime0}({5\over2}^-)\to \Xi_c^{*0}\gamma$         &$0.086^{+0.043}_{-0.034}$    & $44^{+100}_{-~44}$               & $44^{+100}_{-~44}$
\\ \hline
~~\multirow{8}{*}{$[\mathbf{ 6}_F, 1, 0, \rho]$} ~~
&~~\multirow{4}{*}{$\Xi_c^{\prime}({1\over2}^-)$} ~~    &\multirow{4}{*}{$2.88\pm 0.14$}      &\multirow{8}{*}{$3\pm 1$}
&$\Xi_c^{\prime+}({1\over2}^-)\to\Xi_c^{{\prime}+}\gamma$      &$0.14^{+0.07}_{-0.04}$  &$44^{+97}_{-43}$                 &\multirow{2}{*}{$59^{+140}_{-~58}$}
\\
&&&&$\Xi_c^{\prime+}({1\over2}^-)\to \Xi_c^{*+}\gamma$         &$0.082^{+0.036}_{-0.026}$  &$15^{+42}_{-15}$                  &
\\  \cline{5-8}
&&&&$\Xi_c^{\prime0}({1\over2}^-)\to\Xi_c^{{\prime}0}\gamma$    &$0.10^{+0.04}_{-0.03}$ &$25^{+50}_{-25}$                        &\multirow{2}{*}{$47^{+110}_{-~47}$}
\\
&&&&$\Xi_c^{\prime0}({1\over2}^-)\to \Xi_c^{*0}\gamma$         &$0.18^{+0.07}_{-0.05}$  &$22^{+60}_{-22}$                         &
\\ \cline{2-3}\cline{5-8}
&~~\multirow{4}{*}{$\Xi_c^{\prime}({3\over2}^-)$}~~     &\multirow{4}{*}{$2.89\pm 0.14$}
&&$\Xi_c^{\prime+}({3\over2}^-)\to\Xi_c^{\prime+}\gamma$     &$0.24^{+0.10}_{-0.08}$    &$130^{+260}_{-130}$               &\multirow{2}{*}{$190^{+420}_{-190}$}
\\
&&&&$\Xi_c^{\prime+}({3\over2}^-)\to \Xi_c^{*+}\gamma$        &$0.16^{+0.06}_{-0.04}$   &$57^{+160}_{-~57}$                       &
\\  \cline{5-8}
&&&&$\Xi_c^{\prime0}({3\over2}^-)\to\Xi_c^{{\prime}0}\gamma$  &$0.17^{+0.07}_{-0.06}$   &$71^{+140}_{-~70}$                      &\multirow{2}{*}{$130^{+310}_{-130}$}
\\
&&&&$\Xi_c^{\prime0}({3\over2}^-)\to \Xi_c^{*0}\gamma$      &$0.30^{+0.12}_{-0.07}$     &$62^{+170}_{-~62}$                        &
\\ \hline\hline
\end{tabular}}
\label{decayc6X}
\end{center}
\end{table*}

\begin{table*}[ht]
\begin{center}
\renewcommand{\arraystretch}{1.5}
\caption{Radiative decay widths of the $P$-wave charmed baryons $\Omega_c$ belonging to the $SU(3)$ flavor $\mathbf{6}_F$  representation.}
\setlength{\tabcolsep}{0.1mm}{
\begin{tabular}{c| c | c | c | c | c | c |c}
\hline\hline
\multirow{2}{*}{Multiplets}&Baryon & ~~~~~~Mass~~~~~~ & ~~~Difference~~~ & \multirow{2}{*}{~~~~~~Decay channels~~~~~~}  & ~~~~Coupling~~~~ & ~~~~Width~~~~ & ~~~~Total~~~~
\\ & ($j^P$) & ({GeV}) & ({MeV}) & &({$\rm GeV^{-1}$})&({keV})&(keV)
\\ \hline\hline
~~\multirow{2}{*}{$[\mathbf{ 6}_F, 0, 1, \lambda]$}~~ &
~~\multirow{2}{*}{$\Omega_c({1\over2}^-)$}~~     &\multirow{2}{*}{$3.03\pm 0.18$}      &\multirow{2}{*}{-}
&$\Omega_c^0({1\over2}^-)\to\Omega_c^0\gamma$           &$0.37^{+0.20}_{-0.14}$   &$1600^{+4200}_{-1600}$                         &\multirow{2}{*}{$2800^{+8300}_{-2800}$}
\\
&&&&$\Omega_c^0({1\over2}^-)\to \Omega_c^{*0}\gamma$      &$0.31^{+0.16}_{-0.13}$  &$1200^{+4100}_{-1200}$                        &
\\ \hline
\multirow{4}{*}{$[\mathbf{ 6}_F, 1, 1, \lambda]$} &
\multirow{2}{*}{$\Omega_c({1\over2}^-)$}     &\multirow{2}{*}{$3.04\pm 0.10$}      &\multirow{4}{*}{$6\pm 2$}
&$\Omega_c^0({1\over2}^-)\to\Omega_c^0\gamma$          &$0.061^{+0.026}_{-0.024}$   & $48^{+69}_{-41}$                       &\multirow{2}{*}{$65^{+98}_{-56}$}
\\
&&&&$\Omega_c^0({1\over2}^-)\to \Omega_c^{*0}\gamma$     &$0.022^{+0.010}_{-0.010}$  & $17^{+29}_{-15}$                     &
\\ \cline{2-3} \cline{5-8}
&\multirow{2}{*}{$\Omega_c({3\over2}^-)$}&\multirow{2}{*}{$3.07\pm 0.09$} &
&$\Omega_c^0({3\over2}^-)\to\Omega_c^0\gamma$         &$0.043^{+0.018}_{-0.016}$   &$30^{+29}_{-24}$      &\multirow{2}{*}{$46^{+46}_{-38}$}
\\
&&&&$\Omega_c^0({3\over2}^-)\to \Omega_c^{*0}\gamma$   &$0.053^{+0.021}_{-0.020}$   & $16^{+17}_{-14}$                         &
\\ \hline
\multirow{4}{*}{$[\mathbf{ 6}_F, 2, 1, \lambda]$}  &
\multirow{2}{*}{$\Omega_c({3\over2}^-)$}     &\multirow{2}{*}{$3.08\pm 0.19$}      &\multirow{4}{*}{$10\pm 4$}
&$\Omega_c^0({3\over2}^-)\to\Omega_c^0\gamma$      &$0.11^{+0.10}_{-0.06}$    &$52^{+190}_{-~52}$                         &\multirow{2}{*}{$66^{+240}_{-~66}$}
\\
&&&&$\Omega_c^0({3\over2}^-)\to \Omega_c^{*0}\gamma$   &$0.10^{+0.08}_{-0.04}$   &$14^{+49}_{-14}$                     &
\\ \cline{2-3}\cline{5-8}
&\multirow{1}{*}{$\Omega_c({5\over2}^-)$}     &\multirow{1}{*}{$3.14\pm 0.18$}      &
&$\Omega_c^0({5\over2}^-)\to \Omega_c^{*0}\gamma$          &$0.029^{+0.022}_{-0.013}$        &$5.2^{+14.6}_{-~5.2}$    &$5.2^{+14.6}_{-~5.2}$
\\ \hline
\multirow{4}{*}{$[\mathbf{ 6}_F, 1, 0, \rho]$} &
\multirow{2}{*}{$\Omega_c({1\over2}^-)$}     &\multirow{2}{*}{$2.99\pm 0.15$}      &\multirow{4}{*}{$2\pm 1$}
&$\Omega_c^0({1\over2}^-)\to\Omega_c^0\gamma$      &$0.19^{+0.07}_{-0.07}$  &$ 76^{+170}_{-~76}$                        &\multirow{2}{*}{$88^{+210}_{-~88}$ }
\\
&&&&$\Omega_c^0({1\over2}^-)\to \Omega_c^{*0}\gamma$  &$0.076^{+0.029}_{-0.024}$ & $ 12^{+36}_{-12}$                        &
\\ \cline{2-3}\cline{5-8}
&\multirow{2}{*}{$\Omega_c({3\over2}^-)$}     &\multirow{2}{*}{$3.00\pm 0.15$}      &
&$\Omega_c^0({3\over2}^-)\to\Omega_c^0\gamma$     &$0.093^{+0.037}_{-0.029}$   & $ 20^{+43}_{-20}$                     &\multirow{2}{*}{$ 32^{+81}_{-32}$ }
\\
&&&&$\Omega_c^0({3\over2}^-)\to \Omega_c^{*0}\gamma$ &$0.14^{+0.06}_{-0.03}$  & $ 12^{+38}_{-12}$                          &
\\ \hline\hline
\end{tabular}}
\label{decayc6O}
\end{center}
\end{table*}

%
\section{Mixing Effects}
\label{sec:mixing}
%
Heavy quark effective theory provides a reliable framework for studying bottom baryons. However, its applicability to charmed baryons is more limited due to the relatively lighter charm quark mass. Consequently, significant mixing can occur among the three $J^P = 1/2^-$ charmed baryons, as well as among the three $J^P = 3/2^-$ states. In this section we focus on the mixing effects between the $[\mathbf{6}_F, 1, 1, \lambda]$ and $[\mathbf{6}_F, 2, 1, \lambda]$ multiplets. Our analysis follows the methodology of Ref.~\cite{Yang:2020zjl}, which investigates the strong decay properties of $P$-wave charmed baryons belonging to the $SU(3)$ flavor sextet $\mathbf{6}_F$.

We explicitly consider the mixing between the $[{\bf 6}_F, 1, 1, \lambda]$ and $[{\bf 6}_F, 2, 1, \lambda]$ doublets. The physical states are related to the HQET basis states via the following transformation:
\begin{eqnarray}
\left(\begin{array}{c}
|\Sigma_c(3/2^-)\rangle_1\\
|\Sigma_c(3/2^-)\rangle_2
\end{array}\right)
&=&
\left(\begin{array}{cc}
\cos\theta & \sin\theta \\
-\sin\theta & \cos\theta
\end{array}\right)
\\ \nonumber && ~~~~~~~ \times
\left(\begin{array}{c}
|\Sigma_c(3/2^-),1,1,\lambda\rangle\\
|\Sigma_c(3/2^-),2,1,\lambda\rangle
\end{array}\right) \, ,
\\
\left(\begin{array}{c}
|\Xi_c^\prime(3/2^-)\rangle_1\\
|\Xi_c^\prime(3/2^-)\rangle_2
\end{array}\right)
&=&
\left(\begin{array}{cc}
\cos\theta & \sin\theta \\
-\sin\theta & \cos\theta
\end{array}\right)
\\ \nonumber && ~~~~~~~ \times
\left(\begin{array}{c}
|\Xi_c^\prime(3/2^-),1,1,\lambda\rangle\\
|\Xi_c^\prime(3/2^-),2,1,\lambda\rangle
\end{array}\right) \, ,
\\
\left(\begin{array}{c}
|\Omega_c(3/2^-)\rangle_1\\
|\Omega_c(3/2^-)\rangle_2
\end{array}\right)
&=&
\left(\begin{array}{cc}
\cos\theta & \sin\theta \\
-\sin\theta & \cos\theta
\end{array}\right)
\\ \nonumber && ~~~~~~~ \times
\left(\begin{array}{c}
|\Omega_c(3/2^-),1,1,\lambda\rangle\\
|\Omega_c(3/2^-),2,1,\lambda\rangle
\end{array}\right) \, ,
\end{eqnarray}
where $\theta$ is the mixing angle. We adopt $\theta = 37^\circ \pm 5^\circ$, consistent with the value used in Ref.~\cite{Yang:2021lce} for the mixing between the $[\mathbf{6}_F, 1, 1, \lambda]$ and $[\mathbf{6}_F, 2, 1, \lambda]$ doublets. The results are summarized in Table~\ref{tab:result}. 

\begin{table*}[ht]
\begin{center}
\caption{Mass spectra and decay properties of the $P$-wave charmed baryons belonging to the $SU(3)$ flavor $\mathbf{6}_F$ representation, which may be accessible through their radiative decay processes. Possible experimental candidates are listed in the last column for comparison. The strong decay results are taken from Ref.~\cite{Yang:2021lce}, where the subscripts $S$ and $D$ denote $S$-wave and $D$-wave decay channels, respectively.}
\renewcommand{\arraystretch}{1.25}
\scalebox{0.925}{\begin{tabular}{ c |c | c | c | c | c | c | c}
\hline\hline
\multirow{2}{*}{HQET state}&\multirow{2}{*}{Mixing}&\multirow{2}{*}{Mixed state} & Mass & Difference & ~~~~~~~~~~\multirow{2}{*}{Decay channel}~~~~~~~~~~ & Width  & \multirow{2}{*}{Candidate}
\\  &&&   ({GeV}) & ({MeV}) & & ({MeV}) &
\\ \hline\hline
$[\Sigma_c({1\over2}^-),1,1,\lambda]$& -- &$[\Sigma_c({1\over2}^-),1,1,\lambda]$&$2.73^{+0.17}_{-0.18}$&\multirow{4}{*}{$23^{+19}_{-43}$}&
$\begin{array}{l}
\Gamma_S\left(\Sigma_c({1/2}^-)\to \Lambda_c\rho\to\Lambda_c\pi\pi\right)=9.2^{+37.0}_{-~9.2}~{\rm MeV}\\
\Gamma_S\left(\Sigma_c({1/2}^-)\to \Sigma_c\rho\to\Sigma_c\pi\pi\right)=1.2^{+2.1}_{-1.0}~{\rm MeV}\\
\Gamma_S\left(\Sigma_c({1/2}^-)\to\Sigma_c\pi\right)=37^{+60}_{-28}~{\rm MeV}\\
\Gamma\left(\Sigma_c^0({1/2}^-)\to\Sigma_c^0 \gamma\right)=11^{+30}_{-11}~{\rm keV}\\
\Gamma\left(\Sigma_c^0({1/2}^-)\to\Sigma_c^{*0} \gamma\right)=3.5^{+15.0}_{-~3.5}~{\rm keV}
\end{array}$&$48^{+70}_{-29}$ &-
\\ \cline{1-4} \cline{6-8}
$[\Sigma_c({3\over2}^-),1,1,\lambda]$&\multirow{7}{*}{$\theta={37\pm5^\circ}$}&$|\Sigma_c({3\over2}^-)\rangle_1$&$2.75^{+0.17}_{-0.17}$&&
$\begin{array}{l}
\Gamma_D\left(\Sigma_c({3/2}^-)\to\Lambda_c\pi\right)=13^{+20}_{-~9}~\rm MeV\\
\Gamma_D\left(\Sigma_c({3/2}^-)\to\Sigma_c\pi\right)=3.3^{+4.2}_{-2.2}~\rm MeV\\
\Gamma_S\left(\Sigma_c({3/2}^-)\to\Sigma_c^{*}\pi\right)
=6.4^{+10.3}_{-~4.7}~\rm MeV\\
\Gamma({\Sigma^0_c({3/2}^-)\to\Sigma^0_c\gamma})=53^{+130}_{-49}~{\rm keV} \\
\Gamma({\Sigma^0_c({3/2}^-)\to\Sigma^{*0}_c\gamma})=13^{+46}_{-13}~{\rm keV} \\
\end{array}$&$24^{+23}_{-10}$& \multirow{7}{*}{$\Sigma_c(2800)$}
\\ \cline{1-1} \cline{3-7}
$[\Sigma_c({3\over2}^-),2,1,\lambda]$&&$|\Sigma_c({3\over2}^-)\rangle_2$&$2.80^{+0.14}_{-0.12}$&\multirow{4}{*}{$68^{+51}_{-51}$}&
$\begin{array}{l}
\Gamma_D\left(\Sigma_c({3/2}^-)\to\Lambda_c\pi\right)=23^{+35}_{-16}~\rm MeV\\
\Gamma_S\left(\Sigma_c({3/2}^-)\to\Sigma_c^{*}\pi\right)
=3.5^{+6.1}_{-2.7}~\rm MeV\\
\Gamma({\Sigma^0_c({3/2}^-)\to\Sigma^0_c\gamma})=54^{+76}_{-41}~{\rm keV} \\ \Gamma({\Sigma^0_c({3/2}^-)\to\Sigma^{*0}_c\gamma})=7.9^{+15.3}_{-~6.7}~{\rm keV} \\
\end{array}$&$28^{+36}_{-16}$&
\\ \cline{1-4} \cline{6-8}
$[\Sigma_c({5\over2}^-),2,1,\lambda]$&--&$[\Sigma_c({5\over2}^-),2,1,\lambda]$&$2.87^{+0.12}_{-0.11}$&&
$\begin{array}{l}
\Gamma_D\left(\Sigma_c({5/2}^-)\to\Lambda_c\pi\right)=12^{+18}_{-~8}~\rm MeV\\
\Gamma_D\left(\Sigma_c({5/2}^-)\to \Sigma_c\pi\right)=0.39^{+0.72}_{-0.32}~\rm MeV\\
\Gamma_D\left(\Sigma_c({5/2}^-)\to\Sigma_c^{*}\pi\right)
=0.61^{+1.14}_{-0.50}~\rm MeV\\
\Gamma\left(\Sigma_c^0({1/2}^-)\to\Sigma_c^{*0} \gamma\right)=8.5^{+9.2}_{-6.8}~\rm keV\\
\end{array}$&$13^{+18}_{-~8}$&-
\\ \hline
$[\Xi_c^\prime({1\over2}^-),1,1,\lambda]$& -- &$[\Xi_c^\prime({1\over2}^-),1,1,\lambda]$&$2.91^{+0.13}_{-0.12}$&\multirow{5}{*}{$27^{+16}_{-27}$}&
$\begin{array}{l}
\Gamma_S\left(\Xi_c^{\prime}({1/2}^-)\to\Xi_c\rho\to\Xi_c\pi\pi\right)
=1.7^{+7.6}_{-1.7}~\rm MeV\\
\Gamma({\Xi^{0}_c({3/2}^-)\to\Xi^{\prime0}_c\gamma})=35^{+61}_{-30}~{\rm keV} \\
\Gamma_S\left(\Xi_c^{\prime}({1/2}^-)\to \Xi_c^{\prime}\pi\right)=12^{+15}_{-~8}~\rm MeV\\
\Gamma({\Xi^{0}_c({3/2}^-)\to\Xi^{*0}_c\gamma})=13^{+27}_{-13}~{\rm keV} \\
\end{array}$&$14^{+17}_{-~8}$&$\Xi_c(2923)$
\\ \cline{1-4}\cline{6-8}
$[\Xi_c^{\prime}({3\over2}^-),1,1,\lambda]$&\multirow{8}{*}{$\theta={37\pm5^\circ}$}&$|\Xi_c^\prime({3\over2}^-)\rangle_1$&$2.94^{+0.12}_{-0.11}$&&
$\begin{array}{l}
\Gamma_D\left(\Xi_c^{\prime}({3/2}^-)\to \Lambda_c \bar K\right)=2.3^{+4.3}_{-1.7}~\rm MeV\\
\Gamma_D\left(\Xi_c^{\prime}({3/2}^-)\to \Xi_c\pi\right)=4.6^{+8.1}_{-3.3}~\rm MeV\\
\Gamma_D\left(\Xi_c^{\prime}({3/2}^-)\to\Xi_ c^{\prime}\pi\right)=2.0^{+2.2}_{-1.2}~\rm MeV\\
\Gamma_S\left(\Xi_c^{\prime}({3/2}^-)\to \Xi_c^{*}\pi\right)=2.1^{+2.6}_{-1.5}~\rm MeV\\
\Gamma({\Xi^+_c({3/2}^-)\to\Xi^{\prime+}_c\gamma})=52^{+58}_{-33}~{\rm keV} \\
\Gamma({\Xi^+_c({3/2}^-)\to\Xi^{*+}_c\gamma})=19^{+30}_{-15}~{\rm keV} \\
\Gamma({\Xi^{0}_c({3/2}^-)\to\Xi^{\prime0}_c\gamma})=130^{+150}_{-86}~{\rm keV} \\
\Gamma({\Xi^{0}_c({3/2}^-)\to\Xi^{*0}_c\gamma})=41^{+69}_{-30}~{\rm keV} \\
\end{array}$&$12^{+10}_{-~4}$
& $\Xi_c(2939)$
\\ \cline{1-1}\cline{3-8}
$[\Xi_c^\prime({3\over2}^-),2,1,\lambda]$&&$|\Xi_c^\prime({3\over2}^-)\rangle_2$&$2.97^{+0.24}_{-0.15}$&\multirow{4}{*}{$56^{+30}_{-35}$}&
$\begin{array}{l}
\Gamma_D\left(\Xi_c^{\prime}({3/2}^-)\to \Lambda_c \bar K\right)=6.3^{+11.6}_{-~4.7}~\rm MeV\\
\Gamma_D\left(\Xi_c^{\prime}({3/2}^-)\to \Xi_c\pi\right)=11^{+19}_{-~8}~\rm MeV\\
\Gamma_S\left(\Xi_c^{\prime}({3/2}^-)\to \Xi_c^{*}\pi\right)= 1.3^{+1.80}_{-0.94}~\rm MeV\\
\Gamma({\Xi^+_c({3/2}^-)\to\Xi^{\prime+}_c\gamma})=22^{+54}_{-17}~{\rm keV} \\
\Gamma({\Xi^+_c({3/2}^-)\to\Xi^{*+}_c\gamma})=4.6^{+15.5}_{-~4.2}~{\rm keV} \\
\Gamma({\Xi^{0}_c({3/2}^-)\to\Xi^{\prime0}_c\gamma})=63^{+160}_{-~50}~{\rm keV} \\
\Gamma({\Xi^{0}_c({3/2}^-)\to\Xi^{*0}_c\gamma})=12^{+44}_{-12}~{\rm keV} \\
\end{array}$&$19^{+22}_{-~9}$&$\Xi_c(2965)$
\\ \cline{1-4} \cline{6-8}
$[\Xi^\prime_c({5\over2}^-),2,1,\lambda]$&--&$[\Xi^\prime_c({5\over2}^-),2,1,\lambda]$&$3.02^{+0.23}_{-0.14}$&&
$\begin{array}{l}
\Gamma_D\left(\Xi_c^{\prime}({5/2}^-)\to \Lambda_c \bar K\right)=6.3^{+11.4}_{-~4.6}~\rm MeV\\
\Gamma_D\left(\Xi_c^{\prime}({5/2}^-)\to \Xi_c \pi\right)=9.6^{+15.8}_{-~6.8}~\rm MeV\\
\Gamma_D\left(\Xi_c^{\prime}({5/2}^-)\to \Xi_c^{*} \pi\right)=1.5^{+2.6}_{-1.1}~\rm MeV\\
\Gamma({\Xi^{0}_c({3/2}^-)\to\Xi^{*0}_c\gamma})=44^{+100}_{-~44}~{\rm keV}
\end{array}$&$18^{+20}_{-~8}$
&--
\\ \hline
$[\Omega_c({1\over2}^-),1,0,\rho]$& -- &$[\Omega_c({1\over2}^-),1,0,\rho]$&$2.99^{+0.15}_{-0.15}$& \multirow{2}{*}{$12^{+5}_{-5}$} & $\Gamma\left(\Omega_c^0({1/2}^-)\to\Omega_c^0 \gamma\right) =76^{+170}_{-~76}~\rm keV$ &$\neq 0$&\multirow{2}{*}{$\Omega_c(3000)$}
\\ \cline{1-4} \cline{6-7}
$[\Omega_c({3\over2}^-),1,0,\rho]$& -- &$[\Omega_c({3\over2}^-),1,0,\rho]$&$3.00^{+0.15}_{-0.15}$&& $\Gamma\left(\Omega_c^0({3/2}^-)\to\Omega_c^0 \gamma\right) 20^{+43}_{-20}~\rm keV$ &$\neq 0$&
\\ \hline
$[\Omega_c({1\over2}^-),1,1,\lambda]$& -- &$[\Omega_c({1\over2}^-),1,1,\lambda]$&$3.04^{+0.11}_{-0.09}$& \multirow{2}{*}{$27^{+15}_{-23}$} & $\begin{array}{l}
\Gamma\left(\Omega_c^0({1/2}^-)\to\Omega_c^0 \gamma\right)=48^{+69}_{-41}~\rm MeV\\
\Gamma\left(\Omega_c^0({1/2}^-)\to\Omega_c^{*0} \gamma\right)=17^{+29}_{-15}~\rm keV
\end{array}$ &$\neq 0$&$\Omega_c(3050)$
\\ \cline{1-4} \cline{6-8}
$[\Omega_c({3\over2}^-),1,1,\lambda]$&\multirow{4}{*}{$\theta =37\pm5^\circ$}&$|\Omega_c({3\over2}^-)\rangle_1$&$3.06^{+0.10}_{-0.09}$&  &
$\begin{array}{l}
\Gamma_D\left(\Omega_c({3/2}^-)\to \Xi_c \bar K\right)=2.0^{+3.5}_{-1.5}~\rm MeV\\
\Gamma({\Omega^{0}_c({3/2}^-)\to\Omega^{0}_c\gamma})=160^{+140}_{-~86}~{\rm keV} \\
\Gamma({\Omega^{0}_c({3/2}^-)\to\Omega^{*0}_c\gamma})=49^{+62}_{-31}~{\rm keV} \\
\end{array}$&$2.0^{+3.5}_{-1.5}$&$\Omega_c(3066)$
\\ \cline{1-1} \cline{3-8}
$[\Omega_c({3\over2}^-),2,1,\lambda]$&&$|\Omega_c({3\over2}^-)\rangle_2$&$3.09^{+0.22}_{-0.17}$&\multirow{3}{*}{$51^{+26}_{-29}$}&
$\begin{array}{l}
\Gamma_D\left(\Omega_c({3/2}^-)\to \Xi_c \bar K\right)=6.3^{+11.2}_{-~4.8}~\rm MeV\\
\Gamma({\Omega^{0}_c({3/2}^-)\to\Omega^{0}_c\gamma})=18^{+36}_{-16}~{\rm keV} \\
\Gamma({\Omega^{0}_c({3/2}^-)\to\Omega^{*0}_c\gamma})=3.7^{+9.8}_{-3.5}~{\rm keV} \\
\end{array}$&$6.4^{+11.2}_{-~4.8}$&$\Omega_c(3090)$
\\ \cline{1-4} \cline{6-8}
$[\Omega_c({5\over2}^-),2,1,\lambda]$&--&$[\Omega_c({5\over2}^-),2,1,\lambda]$&$3.14^{+0.21}_{-0.15}$&&
$\begin{array}{l}
\Gamma_D\left(\Omega_c({3/2}^-)\to \Xi_c \bar K\right)=5.5^{+9.6}_{-4.1}~\rm MeV\\
\Gamma\left(\Omega_c^0({5/2}^-)\to \Omega_c^{*0} \gamma\right)=5.2^{+14.6}_{-~5.2}~\rm keV
\end{array}$&$5.5^{+9.6}_{-4.1}$&$\Omega_c(3119)$
\\ \hline\hline
\end{tabular}}
\label{tab:result}
\end{center}
\end{table*}

\section{Summary and Discussions}
\label{secsummry}

In this paper we have employed the light-cone sum rule method to systematically investigate the radiative decay properties of the $P$-wave singly charmed baryons in the $SU(3)$ flavor $\mathbf{6}_F$ representation, within the framework of heavy quark effective theory (HQET). The resulting decay amplitudes are summarized separately in Tables~\ref{decayc6S}, \ref{decayc6X}, and \ref{decayc6O} for the $\Sigma_c$, $\Xi_c^\prime$, and $\Omega_c$ baryons, respectively. In addition to the radiative decays, their mass spectra and strong decay properties have been extensively studied in Refs.~\cite{Chen:2015kpa,Chen:2017sci,Yang:2020zjl,Yang:2021lce,Yang:2023fsc}, forming a comprehensive QCD sum rule analysis of $P$-wave singly charmed baryons in the $SU(3)$ flavor $\mathbf{6}_F$ representation under HQET. Notably, for certain baryons with limited strong decay widths, radiative transitions play a crucial role in probing their internal structures and quantum numbers. Furthermore, we have explicitly considered the mixing between the $[\mathbf{6}_F, 1, 1, \lambda]$ and $[\mathbf{6}_F, 2, 1, \lambda]$ doublets. The corresponding results, summarized in Table~\ref{tab:result}, demonstrate that this mixing significantly refines the predicted mass spectra and decay patterns of the associated baryon states.

In particular, compared to the strong decay widths, the radiative decay widths are significant in the following cases:
\begin{itemize}

\item The strong decay widths of the $\Omega_c^0(\frac{1}{2}^-)$ and $\Omega_c^0(\frac{3}{2}^-)$ baryons, which belong to the $[\mathbf{6}_F, 1, 0, \rho]$ doublet, are both calculated to be zero in Ref.~\cite{Yang:2021lce}. In the present study their radiative decay widths are calculated to be $76^{+170}_{-~76}$~keV and $20^{+43}_{-20}$~keV, respectively. The $\Omega_c(3000)^0$ can be interpreted as either of these states, and we propose to confirm this assignment through the $\Omega_c^0\gamma$ decay channel.

\item The strong decay widths of the $\Omega_c^0(\frac{1}{2}^-)$ and $\Omega_c^0(\frac{3}{2}^-)$ baryons, which belong to the $[\mathbf{6}_F, 1, 1, \lambda]$ doublet, are calculated in Ref.~\cite{Yang:2021lce} to be zero and $2.0^{+3.5}_{-1.5}$~MeV, respectively. In the present study their radiative decay widths are calculated to be $48^{+69}_{-41}$~keV and $160^{+140}_{-~88}$~keV, respectively. The $\Omega_c(3050)^0$ and $\Omega_c(3066)^0$ could be interpreted as their experimental candidates, and we propose to confirm these assignments via the $\Omega_c^{(*)0}\gamma$ decay channels.

\end{itemize}
We propose further study of these radiative decay channels in future Belle-II, BESIII, and LHCb experiments.

\section*{Acknowledgments}

This project is supported by
the National Natural Science Foundation of China under Grant No.~12075019,
the National Natural Science Foundation of China under Grants No.~12005172,
the National Natural Science Foundation under contract  No.~12375132,
the Jiangsu Provincial Double-Innovation Program under Grant No.~JSSCRC2021488,
the SEU Innovation Capability Enhancement Plan for Doctoral Students,
the China Postdoctoral Science Foundation under Grants No.~2024M750049,
and
the Fundamental Research Funds for the Central Universities.

\end{document}